\documentclass[prodmode,acmtoms]{acmsmall}
\pdfoutput=1
\usepackage{inconsolata}
\usepackage{amsmath,amssymb}
\usepackage{graphicx}
\graphicspath{{images/}{benchmarks/plots/}}
\usepackage[noend]{algorithmic}

\definecolor{DarkBlue}{rgb}{0.00,0.00,0.55}
\definecolor{DarkRed}{rgb}{0.55,0.00,0.00}
\definecolor{DarkGreen}{rgb}{0.00,0.55,0.00}
\definecolor{Bittersweet}{rgb}{1.0, 0.44, 0.37}
\definecolor{Purple}{rgb}{0.5, 0.0, 0.5}

\usepackage{listings}

\lstdefinelanguage[firedrake]{python}[]{python}{%
  emph={[3]op2,par_loop,READ,WRITE,RW,INC,Set,Map,Dat,DataSet,Global,Const,Mat,Sparsity,Kernel,MixedSet,MixedDat,MixedDataSet,MixedMap,i,Solver},
  emph={[2]Function,Mesh,UnitSquareMesh,FunctionSpace,VectorFunctionSpace,interpolate,Expression,Constant,DirichletBC,solve,assemble,assign,apply},
  emph={dot,div,grad,dx,inner,Coefficient,FiniteElement,VectorElement,TensorElement,TrialFunction,TestFunction,TrialFunctions,TestFunctions,triangle}
}

\lstdefinestyle{framed}{frame=tb}

\lstMakeShortInline{|}

\AtBeginDocument{\setlength{\tempdimen}{\textwidth}}

\newcommand{\Real}{\mathbb{R}}

\acmVolume{0}
\acmNumber{0}
\acmArticle{0}
\acmYear{0}
\acmMonth{0}

\begin{document}

\markboth{F. Rathgeber et al.}{Firedrake: automating the finite element method by composing abstractions.}

\title{Firedrake: automating the finite element method by composing abstractions}

\author{FLORIAN~RATHGEBER, DAVID~A.~HAM, LAWRENCE~MITCHELL, MICHAEL~LANGE, FABIO~LUPORINI, ANDREW~T.~T.~MCRAE, GHEORGHE-TEODOR~BERCEA, GRAHAM~R.~MARKALL and PAUL~H.~J.~KELLY
\affil{Imperial College London}}

\begin{abstract}
  Firedrake is a new tool for automating the numerical solution of partial
  differential equations. Firedrake adopts the
  domain-specific language for the finite element method of the FEniCS project,
  but with a pure Python runtime-only implementation centred on the
  composition of several existing and new abstractions for particular
  aspects of scientific computing. The result is a more complete separation
  of concerns which eases the incorporation of separate contributions from
  computer scientists, numerical analysts and application specialists. These
  contributions may add functionality, or improve performance.

  Firedrake benefits from automatically applying new optimisations. This
  includes factorising mixed function spaces, transforming and vectorising
  inner loops, and intrinsically supporting block matrix operations.
  Importantly, Firedrake presents a simple public API for escaping the UFL
  abstraction. This allows users to implement common operations that fall
  outside pure variational formulations, such as flux-limiters.
\end{abstract}




\acmformat{Florian Rathgeber, David A. Ham, Lawrence Mitchell, Michael
  Lange, Fabio Luporini, Andrew T. T. McRae, Gheorghe-Teodor Bercea, Graham
  R. Markall, and Paul H. J. Kelly, 2015. Firedrake: automating the finite element method by composing abstractions.}

\begin{bottomstuff}
This work was supported by the Engineering and Physical Sciences Research
Council [grant numbers EP/I00677X/1, EP/L000407/1, EP/I012036/1], the Natural
Environment Research Council [grant numbers NE/G523512/1, NE/I021098/1,
NE/K006789/1, NE/K008951/1] and the Grantham Institute, Imperial College
London.  Author's addresses: F.~Rathgeber, (Current address) European Centre
for Medium-range Weather Forecasts; D.~A.~Ham and L.~Mitchell, Department of
Mathematics and Department of Computing, Imperial College London; M.~Lange,
Department of Earth Science and Engineering, Imperial College London;
F.~Luporini G.-T.~Bercea, and P.~H.~J.~Kelly, Department of Computing, Imperial
College London; A.~T.~T.~McRae, (Current address) Department of
Mathematical Sciences, University of Bath; G.~R.~Markall, (Current address)
Embecosm.
\end{bottomstuff}

\maketitle

\section{Introduction}

The numerical solution of partial differential equations (PDEs) is an
indispensable tool in much of modern science and engineering. However, the
successful development and application of advanced PDE solvers on complex
problems requires the combination of diverse skills across mathematics,
scientific computing and low-level code optimisation, which is rarely at
expert level in a single individual. For the finite element method, which
will be the focus of this work, this set of skills includes at least:
knowledge of the system being simulated, analysis of the resulting PDEs,
numerical analysis to create appropriate discretisations, mesh generation,
graph theory to create data structures on those meshes, the analysis and
implementation of linear and nonlinear solvers, parallel algorithms,
vectorisation, and loop nest optimisation under memory constraints.

The development of such software is therefore increasingly a
multi-disciplinary effort and its design must enable scientists with
different specialisations to collaborate effectively without requiring each
one of them to understand every aspect of the system in full detail. The key
to achieving this is to abstract, automate and compose the various processes
involved in numerically solving PDEs. At some level, this process is a
familiar one: few of the people who write C or Fortran code really
understand how the compiler works; and they need not do so. Instead, the
programmer understands the rules of the language and programmes to that
model. Similarly, mathematical operations and results are frequently
employed without having their derivation or proof immediately at hand. In
other words, mathematical and software abstractions such as languages and
theorems enable a separation of concerns between \emph{developing}\ a
technique and \emph{employing} it.

This paper presents a new contribution to the automation and abstraction of
the finite element method. Previous work, most especially the Unified Form
Language \cite{Alnaes2014} employed by the FEniCS project
\cite{book:LoggMardalWells2012,Logg2010} enables scientists to express partial
differential equations (PDEs) in a high productivity interpreted language
close to the mathematics. Implementations of the finite element method have
traditionally been tightly coupled to the numerics, requiring contributors to
have a deep understanding of both. FEniCS creates a separation of concerns
between \emph{employing} the finite element method and \emph{implementing} it.
Firedrake goes beyond this by introducing a new abstraction, PyOP2, to create
a separation within the implementation layer between the \emph{local
discretisation} of mathematical operators, and their \emph{parallel execution}
over the mesh. This separation enables numericists to contribute ever-more
sophisticated finite elements, while computer scientists, expert in parallel
execution but not in numerics, contribute more advanced execution strategies.

In addition to admitting uniformly high performance mesh iteration, the introduction
of the additional abstraction layer results in a very compact code base. The
resulting core Firedrake code has only around 5000 lines of executable code,
while the PyOP2 parallel execution layer has fewer than 9000 executable
lines. This compactness is evidence of the effectiveness of the abstraction
choice made and is of immense benefit to the maintainability and
extensibility of the code base.

Section \ref{sec:abstractions} describes the state of the art in
abstractions for scientific computing, particularly the finite element
method. Section \ref{sec:firedrake_abstractions} details the abstractions,
and their implementations, which are composed to form the Firedrake
toolchain. Sections \ref{sect:PyOP2} and \ref{sec:firedrake} describe in more
detail the Firedrake and PyOP2 abstraction layers which are the core
contribution of this paper. Section \ref{sec:experiments} describes an
extensive computational verification of the performance and capability of
the Firedrake system.

\section{Mathematical and software abstraction of the finite element method}\label{sec:abstractions}

A particular advantage of the finite element method as a class of numerical
methods for PDEs is that the entire algorithm can frequently be described in
highly abstract mathematical terms. In the simplest cases, the mathematics of
the method can be completely specified by a PDE in weak form, along with the
desired boundary conditions and the discrete function spaces from which the
solution and test functions should be drawn. Of course, a complete mathematical
specification of the method is not the same as an efficient, parallel and
bug-free software implementation. As a result, countless years of scientists'
time have been spent over the decades implementing finite element methods in low
level Fortran and C code.

Whilst hand coding algorithms at a low level can produce efficient code, that
approach suffers from a number of serious drawbacks. The key
among these is a premature loss of mathematical abstraction: the symbolic
structure of differential equations, function spaces and integrals is replaced
by loops over arrays of coefficient values and individual floating point
operations. Interspersed among these are parallel communication calls,
threading and vectorisation directives, and so forth.

The original abstract mathematical expression of the equations embodies a
separation of concerns: the equation to be solved is separated from its
discretisation, from the linear and/or non-linear solver techniques to be
applied and from the implementation of the assembly and solvers. A low-level
implementation loses this separation of concerns. This has a number of
deleterious effects. First, choices are committed to far too early: deciding
to change discretisation or the equation to be solved requires the
implementation to be recoded. Second, the developer must deal with the
mixture of equations, discretisation and implementation all at
once. Reasoning about the mathematics of the code requires the developer to
mentally re-interpret series of primitive instructions as the high level
abstract mathematics they represent, and any change made to the desired
behaviour must be implemented by manually working out the correct series of
primitive operations. Changing and debugging the code in this way also carries
the risks of defeating implementation choices which were made to optimise
performance, and of introducing bugs.

\subsection{The benefits and limits of mathematical library interfaces}

Given the limitations of hand-writing low level code, it is unsurprising
that much effort has been devoted to the development of finite element and
other scientific software which maintains something of the mathematical
abstraction of the methods. A core feature of these approaches is that they
present a programming environment in which the data objects correspond to
the higher-level mathematical objects found in the finite element
method. For example there may be data objects corresponding to sparse
matrices, distributed vectors, finite elements and function spaces.

A common and highly successful approach to this is for these mathematical
objects to be represented as data objects in object-oriented
libraries. High level mathematical operations are then expressed as
operations on these objects, resulting in method calls. The actual
implementation of the primitive numerical operations on arrays of floating
point numbers is hidden in the implementation of those
methods. Deal.II \cite{Bangerth2007,Bangerth2013} and
Dune-FEM \cite{Dedner2010} are prominent examples of object-oriented finite element
packages, and there are many others. The object-oriented library approach has
also been very successfully
applied by leading sparse linear algebra libraries, notably including
PETSc \cite{Balay2014} and Trilinos EPetra and TPetra
packages \cite{Heroux2005}.

The library approach is most successful where the mathematical operations
specified by the application developer have a fairly large granularity: for
example in the case of linear algebra packages, the smallest operations
(such as scaling vectors or taking a dot product) still involve an iteration
over the entire vector, and operations such as a linear solve are much
larger. This means that the implementation of tight inner loops and much or
all of the parallelism can be hidden from the application developer, thereby
achieving the desired separation of algorithm and implementation.

Conversely, a key domain of variability in the field of numerical methods
for PDEs lies at the level of the innermost loops: the numerical operations
conducted at each mesh entity (cell, face, edge or vertex) depend on the PDE
being solved and the numerical method employed. This means that the level of
code at which the algorithm is expressed is at or below the level at which
factors such as loop order, data layout and function calls become
performance critical. Fine grained parallelism, such as threading and
vectorisation may also need to be expressed at this level. In contrast to
the case of linear algebra, in the PDE discretisation a critical part of the
user algorithm describes very fine-grain operations which must be woven together
to form an efficient, parallel implementation. For this reason,
library-based finite element packages such as Dune-FEM and Deal.II require
that C++ implementations of integrals expressed as low-level sums
over quadrature points be provided by the application developer.


\subsection{Domain specific languages for finite elements}

The desire to express the integrals at the heart of the finite element
method in a high-level mathematical language while still producing efficient
low-level code implementing these integrals has led some projects to adopt a
different approach. Rather than writing directly executable code utilising
library calls to access functionality, the numerics of the finite element
method are specified purely symbolically in a special purpose language. A
specialised compiler or interpreter then uses this input to generate
low-level, efficient code. Within this category, we can distinguish between
stand-alone languages with their own parser, and embedded languages
implemented in an existing general purpose compiled or interpreted
language. A prominent example of the former class is
FreeFem++ \cite{Hecht2012}, while the Unified Form Language (UFL,
\cite{Alnaes2014}) and Sundance \cite{Long2010} are examples of finite
element domain specific languages (DSLs) embedded in Python and C++
respectively.

A well designed DSL not only enables the application programmer to express
their problem clearly, mathematically and concisely, it also provides the
compiler writer with a great deal of freedom to make optimal implementation
choices, including those which are too verbose, tedious and error-prone to
implement by hand. For example the FEniCS Form Compiler, which takes UFL
as its input language, has been used to develop highly optimised quadrature
\cite{Olgaard2010} and tensor reduction \cite{Kirby2005}\
implementations of finite element assembly.

A further benefit of the DSL approach is that the symbolic mathematical form
of the variational problem is available in the program code. This can be
exploited to automate reasoning about the mathematical structure of the
problem, for example to provide high-level
differentiation of the algorithm with respect to any of its inputs. This is
employed by Sundance and FEniCS \cite{book:LoggMardalWells2012}\ to
compute the linearisation of the terms in the equation. It has been further
exploited to provide automated adjoint operators, and thereby adaptive error
control, functional optimisation and stability analysis
\cite{Rognes2013,Farrell2012a,Funke2013,Farrell2014}.

\section{Exploiting composable abstractions in Firedrake}\label{sec:firedrake_abstractions}

The novel contribution of Firedrake as a piece of mathematical software is
to take the decomposition of the finite element method into automated
abstractions further than previous approaches.  In particular, we
use a uniform abstraction (PyOP2) for the specification of iterations over
the mesh, motivated by the observation that the mathematical statement
of finite element problems decouples the local computation from its
execution over the whole domain. 

\begin{figure}
  \centering
  \includegraphics[width=.65\textwidth]{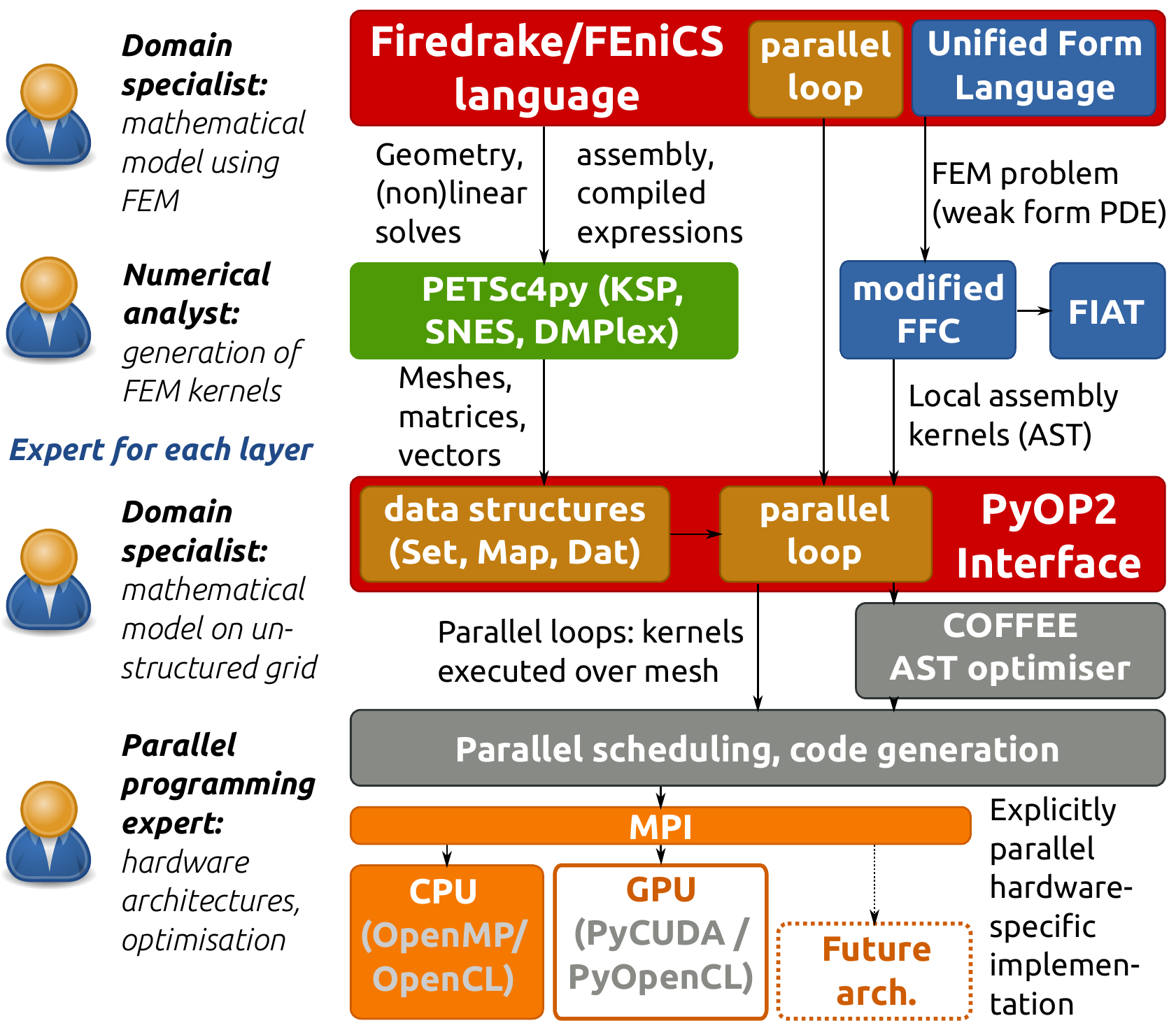}
  \caption{The abstractions composed to create the Firedrake toolchain, and
    the separation of concerns this creates. External tools adopted and/or
    modified from the FEniCS project are in blue, while the tools adopted
    from the PETSc project are in green. Interface layers are represented in
    red while PyOP2 objects are brown. Our code generation
    and execution layer is represented in grey, and the underlying execution
    platform is shown in orange. For the reasons given in
    section \ref{sec:nogpu}, this paper presents results for only the CPU backend.}
  \label{fig:firedrake_toolchain}
\end{figure}

Firedrake models finite element problems as the composition of several
abstract processes and its software stack is composed of separate packages for
each. The core Firedrake package, composes these into a largely seamless
abstraction for finite element problems. Fig.~\ref{fig:firedrake_toolchain}
illustrates the Firedrake software stack, showing the relationships between
the various abstractions and software layers. These are described in more
detail in the following sections.

An important benefit of well-designed mathematical abstractions is that they
facilitate code reuse. Where possible we have adopted and adapted existing
abstractions and existing implementations of those abstractions. This not
only saves re-invention of previous work, it means that users and developers
of those aspects of Firedrake do not need to learn new interfaces. However, in
the case of the tasks of iteration over the mesh graph and the generation of
optimal kernel implementations, there was no completely suitable existing
solution and so new components were created.

\subsection{Specification of finite element problems: \emph{the FEniCS Language}}

The end user of Firedrake wants to specify and solve finite element problems.
In some sense the core part of this is the specification of the weak form of
the PDE, and the selection of the appropriate finite elements. The Unified
Form Language is a particularly elegant and powerful solution to this problem
\cite{Alnaes2014}. UFL is a purely symbolic language with well-defined,
powerful and mathematically consistent semantics embedded in Python. This
makes interactive use possible and allows Firedrake to use the original
implementation of UFL directly, thereby automatically maintaining
compatibility with other users of the language. Firedrake adds several
extensions to UFL, some of which have already been merged back into the
upstream version.

The specification of the PDE and finite elements is necessary but not
sufficient to specify a finite element problem. In addition to the weak form
of the PDE, it is necessary to specify the mesh to be employed, set field
values for initial and/or boundary conditions and forcing functions, and to
specify the sequence in which solves occur. UFL was developed as part of the
FEniCS project, which provides a complete finite element problem solving
environment in the form of the Python interface to DOLFIN \cite{Logg2012}. We
refer to the language for finite element problems defined by DOLFIN and UFL as
the FEniCS language. To ensure compatibility, Firedrake implements (a close
variant of) that language  and presents a user interface which is identical in
most respects to the DOLFIN Python interface.  Firedrake implements various
extensions to the language, and there are a few features of DOLFIN which are
not supported.

\begin{lstlisting}[language={[firedrake]python},style=framed,float,numbers=left,
caption={Firedrake code for the Poisson equation. \lstinline+mesh+ and
\lstinline+degree+ are assumed to have been defined previously. UFL functions
and operations are defined in orange, other FEniCS language constructs in
blue.},label=lst:poisson]
V = FunctionSpace(mesh, "Lagrange", degree)

bc = DirichletBC(V, 0.0, [3, 4])  # Boundary condition for y=0, y=1

u = TrialFunction(V)
v = TestFunction(V)
f = Function(V).interpolate(Expression(
    "48*pi*pi*cos(4*pi*x[0])*sin(4*pi*x[1])*cos(4*pi*x[2])"))
a = inner(grad(u), grad(v))*dx
L = f*v*dx

u = Function(V)
A = assemble(a, bcs=bc)
b = assemble(L)
bc.apply(b)
solve(A, u, b, solver_parameters={'ksp_type': 'cg',
                                  'pc_type': 'hypre',
                                  'pc_hypre_type': 'boomeramg',
                                  'pc_hypre_boomeramg_strong_threshold': 0.75,
                                  'pc_hypre_boomeramg_agg_nl': 2,
                                  'ksp_rtol': 1e-6,
                                  'ksp_atol': 1e-15})
\end{lstlisting}

A Poisson and linear wave equation finite element problem specified in the
FEniCS language for execution by Firedrake are shown in listings
\ref{lst:poisson}\ and \ref{lst:wave} (Section \ref{sec:experiments}).  Line 1
defines a finite element function space on a given mesh (whose definition is
omitted for brevity) and degree using linear Lagrange elements.  A Dirichlet
boundary condition of value 0 on a region of the domain identified by the
markers 3 and 4 is defined on line 3. Lines 5-10 show the UFL code defining
the bilinear and linear forms $a = \nabla u \cdot \nabla v~\mathrm{d}x$ and $L
= fv~\mathrm{d}x$ with test and trial functions $u$ and $v$ and forcing
function $f$. The resemblance to the mathematical formulation is immediately
apparent. In lines 13-15, the forms are assembled into a matrix |A| and
Function |b| with the boundary conditions applied. The linear system of
equations is solved in line 16 for a Function |u| defined on line 12.

\subsection{Finite element tabulation: \emph{FIAT}}

Firedrake employs the FInite element Automatic Tabulator, FIAT
\cite{Kirby2004} which implements the classical finite element abstraction
of \citeN{Ciarlet1978} to support a wide range of finite elements with
relatively few element-specific alterations. The process of merging
Firedrake's extensions to FIAT back into the original version is underway.

\subsection{Iteration over the mesh graph: \emph{PyOP2}}

In a typical finite element problem, the operations whose cost in data
movement or floating point operations is proportional to the size of the mesh
will be the dominant cost.  These operations typically fall into two
categories: iterating over data structures associated with the mesh, and
sparse linear algebra. Firedrake's solution to the former class of operation
is PyOP2 \cite{Rathgeber2012,Markall2013}.

PyOP2 is a domain-specific language embedded in Python for the parallel
execution of computational kernels on unstructured meshes or graphs.
Fundamental concepts are shared with OP2 \cite{Giles2011a}, however the
implementation differs in ways that are crucial for the integration with
Firedrake and other projects. PyOP2 dynamically generates code at runtime by
leveraging Python to inspect objects and data structures. OP2 relies on static
analysis of an input programme, which is transformed through source-to-source
translation at compile time, making it very difficult to embed in another
application. Furthermore, PyOP2 provides sparse matrices and other data
structures required for finite element computations which are not supported by
OP2.

PyOP2 provides an abstract interface for the definition of operations
composed of the application of a kernel function for each entry in a fixed
arity graph. By representing the computational mesh as such a graph, it
becomes possible to represent all of the mesh-visitor operations in the
finite element method as instances of this single abstraction.
A particularly clean separation of concerns is thereby achieved between
the specification of the local kernel functions, in which the numerics of
the method are encoded, and their efficient parallel execution. PyOP2 is the
key novel abstraction in the Firedrake system. It is documented in much
more detail in section \ref{sect:PyOP2}.

\subsection{Unstructured meshes: \emph{DMPlex}}

PyOP2 has no concept of the topological construction of a mesh: it
works with indirection maps between sets of topological entities and
sets of degrees of freedom but has no need to know the origin of these
maps. Firedrake derives the required indirection maps for input meshes
through an intermediate mesh topology object using PETSc's DMPlex API,
a data management abstraction that represents unstructured mesh data as
a directed acyclic graph~\cite{Knepley2009,Balay2014}. This allows
Firedrake to leverage the DMPlex partitioning and data migration
interfaces to perform domain decomposition at runtime while supporting
multiple mesh file formats. Moreover, Firedrake reorders mesh entities
to ensure computational efficiency through communication-computation
overlap, while also employing mesh renumbering techniques provided by
DMPlex to improve cache coherency within the resulting data
sets~\cite{Lange2015}.

\subsection{Linear and non-linear solvers: \emph{PETSc}}

As noted above, the encapsulation of solvers for linear and non-linear systems
of equations is one of the most spectacular success stories for abstraction
in scientific computing. The creation of efficient solver algorithms and
implementations is also a complex and deep research field which it is not
profitable to attempt to reinvent. We therefore adopt the widespread practice
of passing solver problems on to an established high performance solver
library. PETSc is adopted as a particularly well-established and
fully-featured library which provides access to a large range of its own and
third party implementations of solver algorithms \cite{Balay2014}. The
fully featured Python interface to PETSc \cite{Dalcin2011} makes its
integration with Firedrake particularly
straightforward. Employing PETSc for both its solver library and for DMPlex
has the additional advantage that the set of library dependencies required by
Firedrake is kept small.

\section{PyOP2}\label{sect:PyOP2}

Many numerical algorithms and scientific computations on unstructured
meshes can be viewed as the independent application of a local operation
everywhere on the mesh. In the finite element method, this
characterisation applies most obviously to the assembly of integrals over
the domain, however it also applies to other operations such as time step
increments and boundary condition implementation. This local operation is
often called a computational kernel and its independent application lends
itself naturally to parallel computation.

\subsection{Sets}
\label{sec:pyop2:concepts:sets}

A mesh is modelled in PyOP2 as a graph defined by sets of entities (such as
vertices, edges, and cells), and maps between these sets. Sets are used to
represent collections of topological entities: vertices, edges, faces and
cells. Sets are completely abstract entities: they store no bulk data
themselves but only record the number of entities they contain, and their
distribution among MPI processes. A set may also represent a set of nodes at
which data may be stored; this set of nodes need not correspond to a set of
topological entities. This facilitates the support of higher order finite
element spaces in which varying numbers of degrees of freedom may be
associated with various classes of topological entities. Sets exist only to be
the subject of reference of other data objects, most particularly Maps and
Dats.

\subsection{Maps}
\label{sec:pyop2:concepts:maps}

A map associates a tuple of entries in a \emph{target} set with each entry
of another \emph{source} set. For example, the source set might be the set
of cells in a mesh, and the target set might be the set of degrees of
freedom of a finite element function space. The map could then record, for
each cell, the tuple of degrees of freedom of the target function space
which are incident to that cell.

It is important to note that PyOP2 itself has no concept of meshes, or
function spaces. The semantic meanings of sets and maps are defined and
understood only by the Firedrake layer. At the PyOP2 layer these structures
are merely objects over which iteration and indirection can occur.

There is a requirement for the map to be of \emph{constant arity}, that is
each element in the source set must be associated with a constant number of
elements in the target set.
The constant arity restriction causes the extent of many tight loop bounds
to be fixed, which creates opportunities for vectorisation and other
optimisations. However it excludes certain kinds of mappings. A map from
vertices to incident edges or cells is only possible on a very regular mesh
where the multiplicity of any vertex is constant. However the full set of maps
required to implement the finite element method is supported.

\subsection{Data}

PyOP2 supports three core arrangements of mutable data: Dats, which are
abstracted discretised vectors, Mats, which are sparse matrices, and
Globals, which represent data not associated with individual set members. In
other words, a Mat is equivalent to a bilinear operator over a pair of Sets,
a Dat is equivalent to a linear operator over a Set and a Global is a scalar
(a 0-linear operator).

A Dat represents a vector of values, each associated with a particular
member of the Set\footnote{There is actually a thin intermediate Dataset
between the Set and Dat to parametrise the size of the data at each set
element, but this is an implementation detail over which we will not
dwell.} over which that Dat is defined. The Dat presents a completely
abstracted interface: the data may actually reside on one or more
accelerators (GPUs) and be distributed over multiple MPI processes but the
user will not usually observe this. In particular, Dats are able to reason
about the validity and location of their data so that copies to and from the
GPU and halo exchanges over MPI happen automatically and only if required.

A Mat object represents a sparse matrix, that is a linear operator from the
data space defined on one Set to that defined on another. The matrix
interface is actually a fairly thin layer over PETSc (in the CPU case) or
CUSP (in the NVIDIA GPU case) and linear solves are completely outsourced to
those libraries. At this stage, PETSc is the far more complete system and
the only one considered production-ready. The primary role of the Mat object
is to match the sparse linear algebra library abstraction to the PyOP2
abstraction so that a PyOP2 kernel can be employed to assemble matrix
entries efficiently and in parallel.

A Global represents a single tuple of values
not connected with a Set. The reason for including this type,
rather than simply employing a native Python numerical type, is to
facilitate reasoning about updating data location. This enables the PyOP2
system to ensure that a Global always has the correct, consistent value even
when updated in parallel or located on an accelerator.

\subsection{Parloops and kernels}
\label{sec:pyop2:parloop}

PyOP2's model of execution is one of parallel loop operations which
transform the system state, consisting of a set of Dats, Mats and
Globals.  Each parallel loop operation executes a kernel function once
for each member of a specified \emph{iteration set}.  In finite
element computations, this set is usually the set of a particular
class of topological entities, thereby allowing a stencil operation to
be executed over the whole mesh. The function usually accesses each
Dat argument $f$ indirectly through a map $m$. In other words, when a
kernel function $k$ is called for iteration set entry $e$, then the
reference to the set of values given by $f(m(e))$ is passed to $k$.
For a computation over cells where $k$ requires data $f$ defined over
vertices, $m$ provides the indices into $f$ for each cell $e$.

For example, if $e$ is the set of cells in a mesh, $f$ is the set of degree
of freedom values of a discretised field, and $m$ is the map from cells to
the incident degrees of freedom, then $f(m(e))$ will be a reference to the
set of degree of freedom values incident to $e$.

We term the application of a kernel to a particular set of data a
\emph{Parloop}.  Specification of a Parloop requires a kernel function
$k$, a set of iteration entities $E$, and data arguments $f_i(a_i,
m_i)$ each of which is annotated with an access descriptor $a_i$ and
indirection map $m_i$.  A Parloop created with arguments $(k, E,
f_0(a_0, m_0), \ldots, f_n(a_n, m_n))$ encodes the mathematical
algorithm
\begin{algorithmic}
  \FORALL {$e \in E$}
      \STATE {$k\bigg(f_{0}(m_0(e))\ldots, f_{n}(m_{n}(e))\bigg)$},
  \ENDFOR
\end{algorithmic}
where each element $m_i(e)$ of $f_i$ is accessed according to the
descriptor $a_i$ as detailed in the next subsection.

The kernel only has access to those entries of the Dat arguments
which are adjacent to the current iteration set entry under the map
provided. It sees the local ordering of the Dat entries to
which it has access, but has no information about the global
indices.

The loop over the iteration set $E$ is explicitly unordered and parallel:
PyOP2 is licensed to execute it in any order and using as many threads, vector
lanes or distributed processes as are available.  Indirect access to data
creates the possibility that this parallel execution may cause write
contention, that is the same piece of data is accessed via more than one
entity in the iteration set. PyOP2 must reason to avoid these contentions
using colouring, communication and copies of data as appropriate. This is made
possible by the specification of access descriptors for all kernel
arguments.

The current colouring implementation in PyOP2 is deterministic, which
results in bit-reproducible results when run on the same number of
processors. Whether this feature remains sustainable as hardware parallelism
becomes more fine-grained is yet to be determined.

\subsubsection{Access descriptors}

Kernel functions modify their data arguments in place. The critical
observation in OP2, which is adopted in PyOP2, is that mesh-based simulation
kernels modify their arguments in characteristic ways. By explicitly
specifying the character of the operations which the kernel will perform on
each Dat, automated reasoning about the parallel execution of the Parloop in
the presence of indirectly accessed arguments becomes vastly easier.
The available access descriptor are as follows:
\begin{description}
\item[READ] The kernel may use previous values of this argument but not set
  them.
\item[WRITE] The kernel may set the values of this argument, but the kernel's
  behaviour does not depend on the previous values of the argument.
\item[RW] The kernel may set the values of this argument and may use the
  previous values of the argument. Note that this still does not imply a
  particular execution order over the iteration set.
\item[INC] The kernel adds increments to the values of the argument using
  the equivalent of the \verb-+=- operator in C.
\end{description}
The reader will immediately observe that READ, WRITE, and INC are special
cases of RW. However their inclusion enables more sophisticated automated reasoning
about data dependencies than would be possible were all arguments labelled RW.

Any data accessed as READ, RW or INC is automatically gathered via the mapping
relationship in a \emph{staging in} phase and the kernel is passed pointers to
local data. After the kernel has been invoked, any data accessed as WRITE, RW
or INC is scattered back out in a \emph{staging out} phase. Only data accessed
in INC mode could potentially cause conflicting writes and requires thread
colouring to prevent any contention.

\subsubsection{Global arguments}

Global reductions are important operations in mesh-based simulations. Users
may wish to calculate globally integrated quantities, such as energy, or
execute other reductions such as calculating the maximum Courant number in
the simulation domain. Global data does not have an indirection map
relationship with the mesh: the same global value is visible from every mesh
entity. The kernel is therefore passed an appropriately sized variable into
which it can place its contribution to the reduction
operation. Globals have their own set of permitted access descriptors which
reflect this: READ, SUM, MIN, MAX. PyOP2 is free to create multiple
variables in memory corresponding to a single Global to support
parallel kernel execution. The access descriptor enables PyOP2 to
subsequently reduce these multiple variables to a single value. The addition
of further reduction access descriptor operations, or even allowing
user-specified reductions, would be straightforward. However at the time of
writing there does not appear to be user demand for this feature.

\subsubsection{Matrix arguments}

Mat arguments differ from Dat and Global arguments in a number of important
ways. Critically, from PyOP2's perspective, Mats are write-only data
structures. Operations which read matrices, such as matrix-vector multiply
and solving linear systems, are executed by the sparse matrix library (for
CPU execution this is PETSc). Consequently, the only access descriptors
permitted for Mats are WRITE and INC. A Mat represents a linear relationship
between two sets, corresponding to the rows and the columns of the matrix,
so two maps (which may be identical) are required to map the kernel
contribution to the matrix. In terms which may be more familiar to the reader
conversant with the finite element method, the kernel is responsible for the
local assembly of the integral of a test function against a trial function,
and PyOP2 then uses the Maps to execute the global assembly into a sparse
matrix.

\subsection{Kernel optimisation in COFFEE}

Kernels are initialised with either a C code string or an abstract syntax tree
(AST), from which C code is generated. The AST representation provides the
opportunity for optimisation through the COFFEE AST optimiser
\cite{Luporini2015}, a compiler which specialises in advanced optimisations
for short loops enclosing non-trivial mathematical expressions of the kind
which typify finite element local assembly kernels.

COFFEE performs platform-specific optimisations on the AST with the goals of
minimising the number of floating-point operations and improving instruction
level parallelism through the use of SIMD (Single Instruction Multiple Data)
vectorisation. The optimiser can detect invariant subexpressions and hoist
them out of the loop nest, permute and unroll loop nests and vectorise
expressions. The last step may require padding of the data and enforcing
alignment constraints to match the target SIMD architecture. COFFEE supports
both SSE (Streaming SIMD Extensions) and AVX (Advanced Vector Extensions)
instruction sets.

\section{The Firedrake layer}\label{sec:firedrake}

The role of the Firedrake layer is to marshal the abstractions provided by
UFL, FIAT, FFC, PETSc and PyOP2 to take finite element problems specified in
the FEniCS Language and efficiently produce solutions.

\subsection{Mapping finite element constructs to data abstractions}

\begin{figure}[ht]
\centering
\includegraphics[width=.6\textwidth]{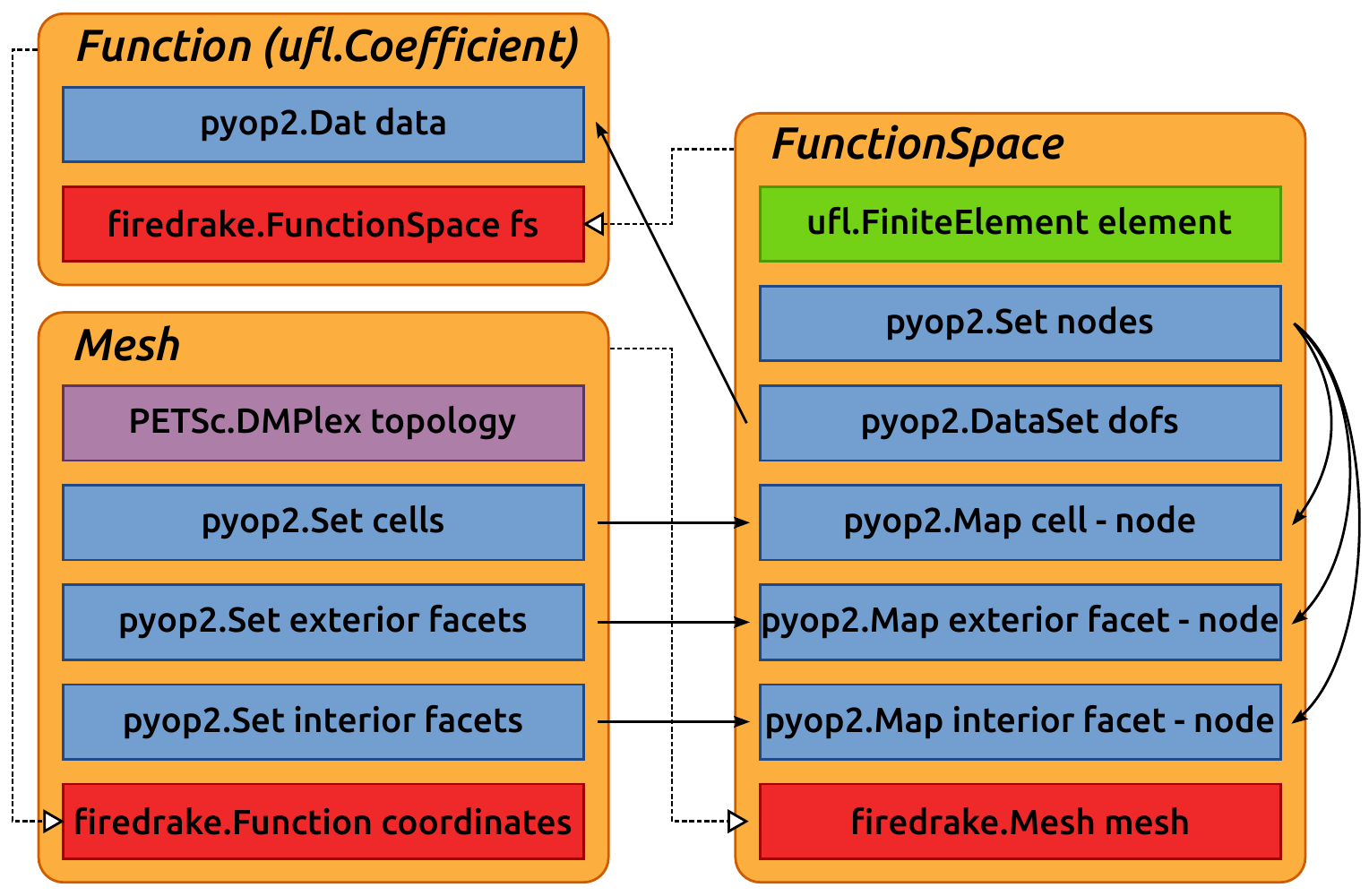}
\caption{The PyOP2 and PETSc objects of which key Firedrake objects are
  composed. PyOP2 objects are shown in blue, references to other Firedrake
  objects in red and PETSc objects are purple.}
\label{img:firedrake_types}
\end{figure}

The FEniCS Language presents higher level mathematical objects than
PyOP2. Firedrake implements these by composing suitable combinations of PyOP2
and PETSc objects. Fig.~\ref{img:firedrake_types} illustrates this
relationship. The Firedrake implementation of operations in the FEniCS
language consists primarily of selecting the relevant PyOP2 objects and
composing corresponding parallel loop calls so that the PyOP2 layer can
undertake the actual calculation.

\subsubsection{The mesh abstraction}

The primary functions of the mesh object are to record adjacency between the
topological entities (vertices, edges, faces, facets and cells) of the mesh,
and to record the mesh geometry. The former of these is encoded in a PETSc
DMPlex which provides arbitrary adjacency relationships \cite{Knepley2009}.

A common approach in PDE toolkits is to treat the coordinates as a special
class of data by, for example, storing the coordinates of each vertex in the
mesh. Firedrake eschews this approach in favour of treating the coordinates as
a first class vector-valued field represented in a suitable vector-valued
function space. An advantage of this approach is that any operation which can
be applied to a field may be applied to the coordinates. Solving a finite
element problem to determine a new geometry field is therefore straightforward.
Representing the coordinates using a fully featured function space also
presents a mechanism for supporting curved (isoparametric) elements, support
for which is currently available in a branch of Firedrake.

The mesh contains PyOP2 Sets which are proxies for the sets of cells,
interior and exterior facets\footnote{A facet is a mesh entity of
  codimension 1: an edge of a 2D mesh or a face of a 3D mesh.}  of the
mesh. These form the iteration spaces for the PyOP2 parallel loops and,
correspondingly, are the ``source'' sets of Maps which encode function
spaces.

\subsubsection{Function spaces and functions}

A central and distinct feature of the finite element method is its
representation of all solution fields as weighted sums of basis
functions. The FEniCS language supports this construction with
FunctionSpace and Function objects. The Function objects store the
coefficient values and a reference to the corresponding FunctionSpace,
while the FunctionSpace stores all of the indirections from the Mesh
to the degrees of freedom, and the symbolic information required to
access the basis functions. As Fig.~\ref{img:firedrake_types}\
demonstrates, this maps in a rather natural way onto PyOP2 data types.

The Function object holds a PyOP2 Dat. This reflects the separation of
concerns in the Firedrake toolchain: the Firedrake layer reasons about the
finite element method and all of the actual data storage and communication
is delegated to PyOP2.

FunctionSpace objects contain PyOP2 Map objects which encode the
indirections from mesh entities to the collections of degrees of freedom
required to implement the finite element method. The cell-node indirection
map provides the indices of the nodes incident to each cell. Equivalently,
this is the set of nodes whose associated basis functions may be non-zero in
that cell. The use of the term \emph{node}\ here rather than \emph{DOF}
(degree of freedom) reflects the treatment of vector and tensor function
spaces: the same indirection map is employed regardless of the number of
degrees of freedom at each mesh location, and the indices of the DOFs are
calculated from this.

\subsubsection{Assembling forms}

Solving a variational problem requires the assembly of a linear system of equations
in the linear, and the Jacobian and residual form in the non-linear case.  In
the Poisson problem in listing \ref{lst:poisson}, the bilinear and linear
forms |a| and |L| are explicitly assembled into the sparse matrix |A| and
vector |b| respectively. Firedrake hands off assembly computations to PyOP2
Parloops (Section \ref{sec:pyop2:parloop}) with a form-specific list of
arguments constructed as follows: The local assembly kernels for the forms are
generated by FFC as described in the following section.  The iteration set is
extracted from the FunctionSpace of the test function |v| and the first
Parloop argument is the output tensor. For the bilinear form this is a Mat
built from a pair of maps extracted from test and trial space, for the linear
form a Dat obtained by creating a new Function on the test space. The second
Parloop argument is the coordinate field. Each coefficients used in the form,
such as |f| in listing \ref{lst:poisson}, translates into an additional
argument.

\subsection{A modified FFC}

The FEniCS project provides the FEniCS Form Compiler (FFC), which takes
variational forms specified in UFL and generates optimised C++ kernel
functions conforming to the UFC interface \cite{chap:LoggOlgaardEtAl2012}.
This approach cuts across the abstraction provided by the PyOP2 interface: in
PyOP2 the specification of the kernel is a problem-specific question delegated
to the user (in this case, the PyOP2 user is Firedrake). Conversely, the
optimisation of the kernel body for a given hardware platform is a matter for
which PyOP2 (specifically COFFEE) takes responsibility. To reflect this, the
version of FFC employed in the Firedrake toolchain is substantially modified.
It still accepts UFL input but produces an un-optimised (and, indeed,
unscheduled) abstract syntax tree (AST) for the local assembly of the form.
Firedrake employs this AST to create a PyOP2 Kernel, and executes a PyOP2
parallel loop to perform global assembly. The modifications required to FFC
are such that the Firedrake version of FFC is effectively a fork and will
not be merged back. However it is hoped that the UFLACS form compiler,
currently under development by Martin Aln{\ae}s will provide a basis for a
unified compiler infrastructure.

\subsection{Escaping the abstraction}\label{sec:escape}

It is an inherent feature of software abstractions that they create a
division between those algorithms which are expressible in the
abstraction, and those which are not. In a well-designed abstraction,
the former are concise, expressive, and computationally
efficient. However any part of an algorithm not expressible within the
abstraction may become impossible to express without completely
breaking out of the abstract framework and coding at a much lower
level.  It will never be possible to represent all algorithms with the
same level of elegance in a single abstraction. Instead, the challenge
is to ensure that a graceful degradation of abstraction occurs. That
is to say, operations which lie a little outside the abstraction
should require the user to work at only a slightly lower level, and
access to aspects of the abstraction which are still applicable should
be preserved.

\subsubsection{Custom kernels}

The FEniCS language presents an
elegant and powerful abstraction for the expression of the core of the
finite element method: weak form PDEs and their solution on piecewise
polynomial triangulations of domains. However, it is frequently the case
that real simulation challenges also incorporate non-finite element
aspects. For example discontinuous Galerkin discretisations may require
shock detectors and slope limiters; parametrisations of unresolved phenomena
may require complex pointwise operations; and initial conditions may require
access to external data in ways not representable in UFL.

The critical observation is that these operations, and many others besides,
are still characterised by visiting mesh entities and accessing only data
local to them: the operations supported by PyOP2. Firedrake therefore
presents the user with the option of specifying a custom kernel in either C
or as an AST. This kernel can then be explicitly executed over the mesh by
invoking a parallel loop. If, as is often the case, the data access patterns
are equivalent to those of the finite element method, then the user can
invoke the Firedrake wrapper of a parallel loop and let Firedrake extract
the correct Maps and Dats from the Firedrake Function. Alternatively, the
user may directly invoke a PyOP2 parallel loop and extract the PyOP2 data
structures manually. In either case, the automated parallelisation provided
by PyOP2 remains. Listings \ref{lst:kernel}\ and \ref{lst:pyop2kernel} show
an example of a randomised initial condition specified with custom Firedrake
and PyOP2 kernels respectively.

\subsubsection{Direct access to data structures}

At a more direct level, the user may also elect to directly access the data in
the Firedrake data structures. Since Firedrake is a pure Python library, the
user can then deploy the full armoury of Python, NumPy and compatible
libraries. PyOP2 employs the introspection capabilities of Python so that even
in this case it remains aware of the data which has been accessed and
modified. PyOP2 ensures that copies and halo exchanges occur as necessary to
ensure that the user's view of the data is current and correct, and that
algorithmic correctness is maintained.

\subsubsection{Access to generated code}

For debugging purposes, it is sometimes useful for the user to access the C
code which PyOP2 generates. This is accessible both in the disk cache and in
memory attached to the relevant PyOP2 parallel loop object. In the
particular case of C code which fails to compile (most commonly due to a
syntax error in user-provided custom kernel code), the error message
provides the location of the generated source file and the compiler error
log.

\subsection{Additional features facilitated by the Firedrake abstraction}

\subsubsection{Factorisation of mixed function spaces}\label{sec:mixed}

When solving PDEs with multiple unknown solution fields, the standard finite
element approach is to seek a solution in the mixed function space given by
concatenating the function spaces of the solution fields. The test function
is, naturally, drawn from the same space. UFL represents a form defined over
a mixed function space as a single form. FFC then constructs a single mixed
kernel which iterates over the combined set of basis function of the test
space and (in the case of a bilinear form) the trial space. In DOLFIN this
is then assembled into a single monolithic sparse matrix.

In contrast, Firedrake takes the UFL form, represented as an abstract syntax
tree, and employs symbolic manipulation to split it into forms for each
combination of constituent test and trial space. This results in separate
forms for each block of the mixed system, and FFC then creates kernels for
those individual blocks. The resulting kernels have simpler loop structures,
which aids COFFEE in producing highly optimised implementations. Bilinear
forms are then assembled into a hierarchical matrix structure, comprising a
matrix for each block combined using PETSc's nested matrix
facility \cite[p86]{Balay2014}. Using PETSc's compressed sparse row
storage, the insertion of entries into submatrices is expected to be faster
than into a monolithic matrix due to the smaller number of non-zero columns (which
have to be searched) in each row. This furthermore enables more efficient
exploitation of block solver techniques such as Schur complements. A
simulation employing mixed function spaces is presented in section
\ref{sec:cahnhilliard}. A much more detailed exposition of the mixed form
splitting algorithm is presented in \citeN[section 5.2.3]{Rathgeber2014}.

\subsubsection{Pointwise operations}

Users often need to change the values of fields by means other than solving
a variational problem. For example when employing a Runge-Kutta timestepping
scheme, variational problems are solved for the updates to fields, but the
actual updates are linear combinations of fields. Similarly users frequently
choose to calculate forcing functions pointwise in terms of other functions
or may rescale the coordinate field. All
of these are achievable by writing custom kernels, however they are
expressed much more naturally by writing assignments of expressions in which
the variables are Function objects. These expressions are then compiled to
form a kernel function which is applied pointwise over the mesh. The
explicit wave equation code shown in listing \ref{lst:wave}\ illustrates the
simplicity of the user code required. The increments for $p$ and $\psi$ both
employ the pointwise expression compiler.

\subsubsection{Immersed manifolds and extruded meshes}

The support for domains which are manifolds immersed in a higher dimensional
spaces introduced in \citeN{Rognes2013a}\ extends directly to
Firedrake. Furthermore Firedrake has extended the algebraic representation
of finite elements and basis functions in UFL, FFC and FIAT to enable the
algorithmic creation of tensor product finite elements on quadrilateral,
triangular prism, and hexahedral cells \cite{McRae2015}. A particularly
important class of meshes in high aspect ratio domains, such as the
atmosphere and ocean, is composed of layers of triangular prism or
hexahedral cells aligned in the vertical direction. The PyOP2 abstraction
has been extended to exploit the structure induced by this vertical
alignment to create very high speed iteration over such ``extruded''
meshes documented in~\citeN{Bercea2016}.

\section{Experiments}\label{sec:experiments}

Firedrake is a tool chain capable of solving a wide range of finite element
problems, which is demonstrated in this section through experiments chosen to
cover different characteristics of its implementation. These include
assembling and solving a stationary Poisson problem, the non-linear
time-dependent Cahn-Hilliard equation and the linear wave equation using an
explicit time stepping scheme.  Implementation aspects investigated are
assembly of left- and right-hand sides for regular and mixed forms, solving
linear and non-linear systems, evaluating expressions and using fieldsplit
preconditioners.  All benchmarks represent real-world applications used in
fluid dynamics to model diffusion, phase separation of binary fluids and wave
propagation.

The principle contribution of this paper is to describe the composition of
abstractions and consequent separation of concerns achieved in Firedrake. A
comprehensive performance evaluation is beyond its scope, indeed a
comprehensive performance evaluation of a single problem might easily occupy
an entire paper. Instead, this section is designed to enable the reader to
develop an impression of the broad performance characteristics of
Firedrake.

We have chosen to compare against DOLFIN for two reasons. The
first is that it is the package which provides the closest analogue to
Firedrake - many of the same test cases can be run from nearly the
same code. The second reason goes to the heart of the difficulty of
conducting fair performance comparisons. By using someone else's code,
it is difficult to avoid the risk that any performance deficiency is
due to inexpert use rather than an inherent flaw. By employing DOLFIN on
ARCHER using the compilation flags recommended by the DOLFIN developers and
using test cases based on DOLFIN examples, we minimize the chance that any
performance deficiencies are due to incorrect use of  the software.

Source code for all benchmarks and the scripts used to drive them are
available as part of the firedrake-bench repository hosted on GitHub. The
particular version used in these experiments has been archived on Zenodo \cite{firedrake_2016_56646}.

\subsection{Experimental setup}

Computational experiments were conducted on the UK national supercomputer
ARCHER, a Cray XC30 architecture \cite{andersson_cray_2014} with an Aries
interconnect in Dragonfly topology. Compute nodes contain two 2.7 GHz, 12-core
E5-2697 v2 (Ivy Bridge) series Intel Xeon processors linked via a Quick Path
Interconnect (QPI) and 64GB of 1833MHz DDR3 memory accessed via 8 channels and
shared between the processors in two 32GB NUMA regions. Each node is connected
to the Aries router via a PCI-e 3.0 link. For the reasons given in
section \ref{sec:nogpu}, execution is always one core per MPI process:
OpenMP is not employed.

Firedrake and PETSc were compiled with version 4.9.2 of the GNU
compilers\footnote{Due to technical limitations in accessing the licence
  server, Intel and Cray compilers cannot be used on ARCHER compute nodes
  and are therefore unavailable to PyOP2's just in time compilation system.}
and Cray MPICH2 7.1.1 with the asynchronous progress feature enabled was
used for parallel runs.  The Firedrake component revisions used are archived
on Zenodo and are accessible via the DOIs in the relevant citation:
Firedrake \cite{lawrence_mitchell_2016_56640}, PyOP2
\cite{florian_rathgeber_2016_56635}, FIAT
\cite{andrew_t_t_mcrae_2016_56637}, COFFEE \cite{fabio_luporini_2016_56636},
ffc \cite{anders_logg_2016_56643}, PETSc \cite{barry_smith_2016_56641},
PETSc4py \cite{lisandro_dalcin_2016_56639}. The DOLFIN used as a comparator
is revision 5ec6384 (July 12 2015) and is linked to the same PETSc version
as Firedrake.

Generated code is compiled with |-O3 -fno-tree-vectorize| in the Firedrake
and |-O3 -ffast-math -march=native| (as suggested by the FEniCS developers) in
the DOLFIN case.

Unless otherwise noted, DOLFIN is configured to use quadrature representation
with full FFC optimisations and compiler optimisations enabled and Firedrake
makes use of COFFEE's loop-invariant code motion, alignment and padding
optimisations described in \citeN{Luporini2015} using quadrature
representation.  Meshes are reordered using PETSc's implementation of reverse
Cuthill-McKee in the Firedrake case and DOLFIN's mesh reordering respectively.

Benchmark runs were executed with exclusive access to compute nodes and
process pinning was used. All measurements were taken preceded by a dry run of
the same problem to pre-populate the caches for kernels and generated code to
ensure compilation times do not distort measurements. Reported timings are the
minimum of three consecutive runs.

\subsection{Poisson}

Poisson's equation is a simple elliptic partial differential equation. A
primal Poisson problem for a domain $\Omega \in \Real^n$ with boundary
$\partial \Omega = \Gamma_{D} \cup \Gamma_{N}$ is defined as:
\begin{align}
   - \nabla^{2} u &= f \quad {\rm in} \ \Omega, \\
                u &= 0 \quad {\rm on} \ \Gamma_{D}, \\
 \nabla u \cdot n &= 0 \quad {\rm on} \ \Gamma_{N}.
  \label{eq:experiments:poisson:strong}
\end{align}
The weak formulation reads: find $u \in V$ such that
\begin{equation}
  \int_\Omega \nabla u \cdot \nabla v \ \mathrm{d}x = \int_\Omega fv \
  \mathrm{d}x \quad \forall v \in V
  \label{eq:experiments:poisson:weak}
\end{equation}
where $V$ is a suitable function space satisfying the Dirichlet boundary
condition $u = 0 \ {\rm on} \ \Gamma_D$.

This benchmark demonstrates assembly of a bilinear and linear form into a
sparse matrix and vector, and solving a linear system with a preconditioned
Krylov method.

\subsubsection{Problem Setup}

The domain $\Omega$ is chosen to be the unit cube $[0, 1]^3$, represented as a
fully unstructured mesh. The source term $f$ is:
\begin{align}
  f(x, y, z) = 48 \pi^2 \cos(4\pi x)\sin(4\pi y)\cos(4\pi z)
  \label{eq:experiments:poisson:f}
\end{align}
with known analytical solution
\begin{align}
  u(x, y, z) = \cos(4\pi x)\sin(4\pi y)\cos(4\pi z).
  \label{eq:experiments:poisson:u}
\end{align}

Since the operator is symmetric positive definite, the problem is solved using
a CG solver \cite{Hestenes1952} with the HYPRE BoomerAMG algebraic multigrid
preconditioner \cite{Falgout2006} on a unit cube mesh of varying resolution
and for varying polynomial degrees. Listing \ref{lst:poisson} shows the
Firedrake code for this problem.

\subsubsection{Results}

\begin{figure}[htbp]
\centering
\includegraphics[width=\textwidth]{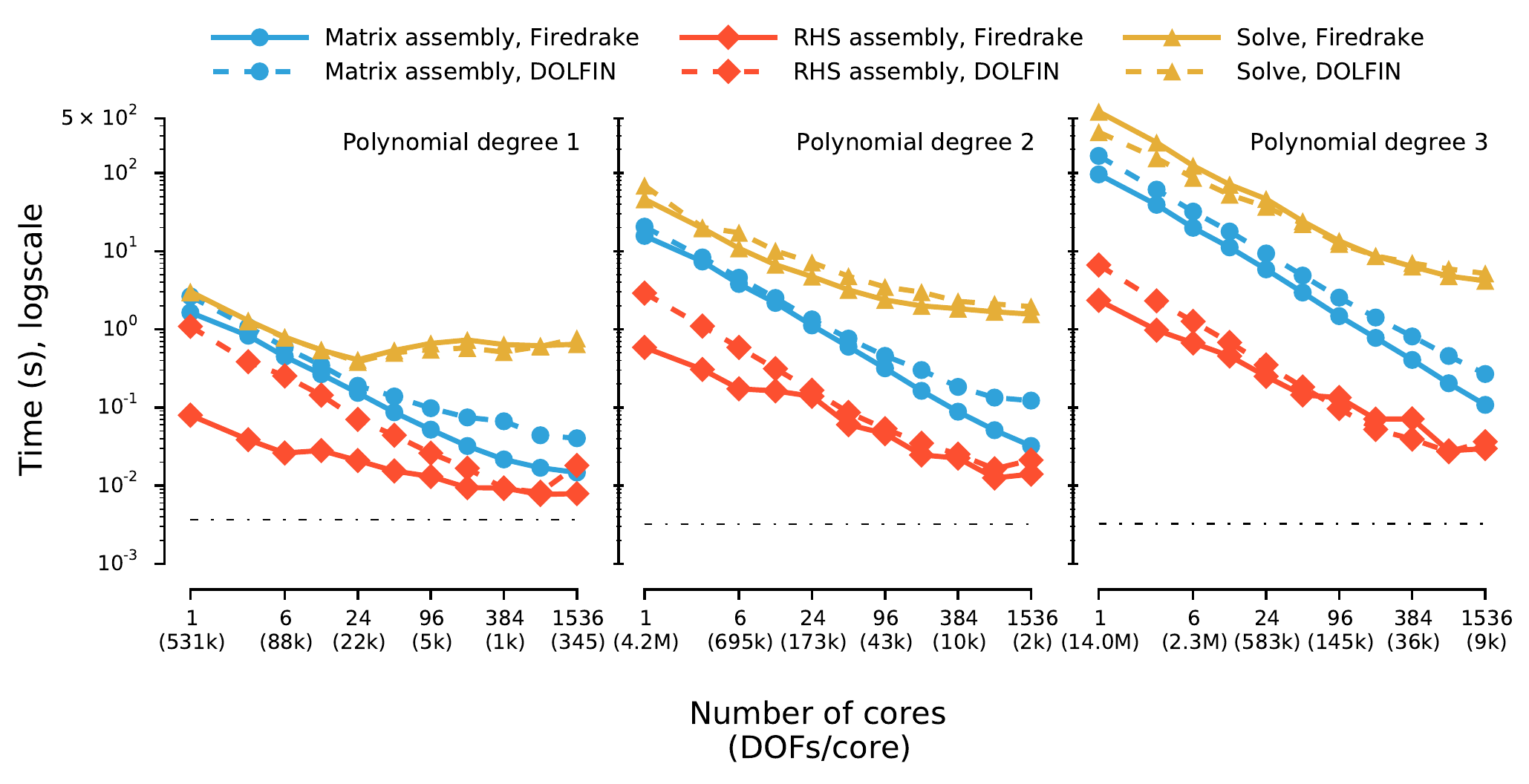}
\caption{Poisson strong scaling for degree one (left),
two (centre) and three (right) basis functions. Overhead for
right hand side assembly is indicated by the horizontal dash-dotted line.
Note that times are log scale.

Solve time clearly dominates in all cases, in particular for higher order and
in the strong scaling limit, where the scaling flattens out at around 10k DOFs
per core.
Firedrake is faster at assembling left- and right-hand sides in almost all cases,
demonstrating the efficiency of low overhead assembly kernel execution through
PyOP2. Matrix assembly is considerably faster in the strong scaling limit in
particular for low order, which can be attributed to Firedrake's way of
enforcing strong boundary conditions described in \citeN[Section
5.5]{Rathgeber2014}.
Right-hand side assembly has a considerably faster sequential base line for
Firedrake such that it is affected by non-parallelisable overheads in the
strong scaling limit sooner than DOLFIN. The sequential overhead indicated for
Firedrake in Fig.~\ref{img:experiments:poisson:strong} causes the scaling to
flatten out much earlier than for matrix assembly. The time spent on
right-hand side assembly however is negligible such that the overall run time
is not greatly affected.
}
\label{img:experiments:poisson:strong}
\end{figure}

Strong scaling runtimes for matrix and right-hand side assembly and linear
solve comparing DOLFIN and Firedrake on up to 1536 cores are shown in
Fig.~\ref{img:experiments:poisson:strong} for problems of approximately 0.5M
to 14M DOFs for first and third order respectively.

\begin{figure}[htbp]
\centering
\includegraphics[width=\textwidth]{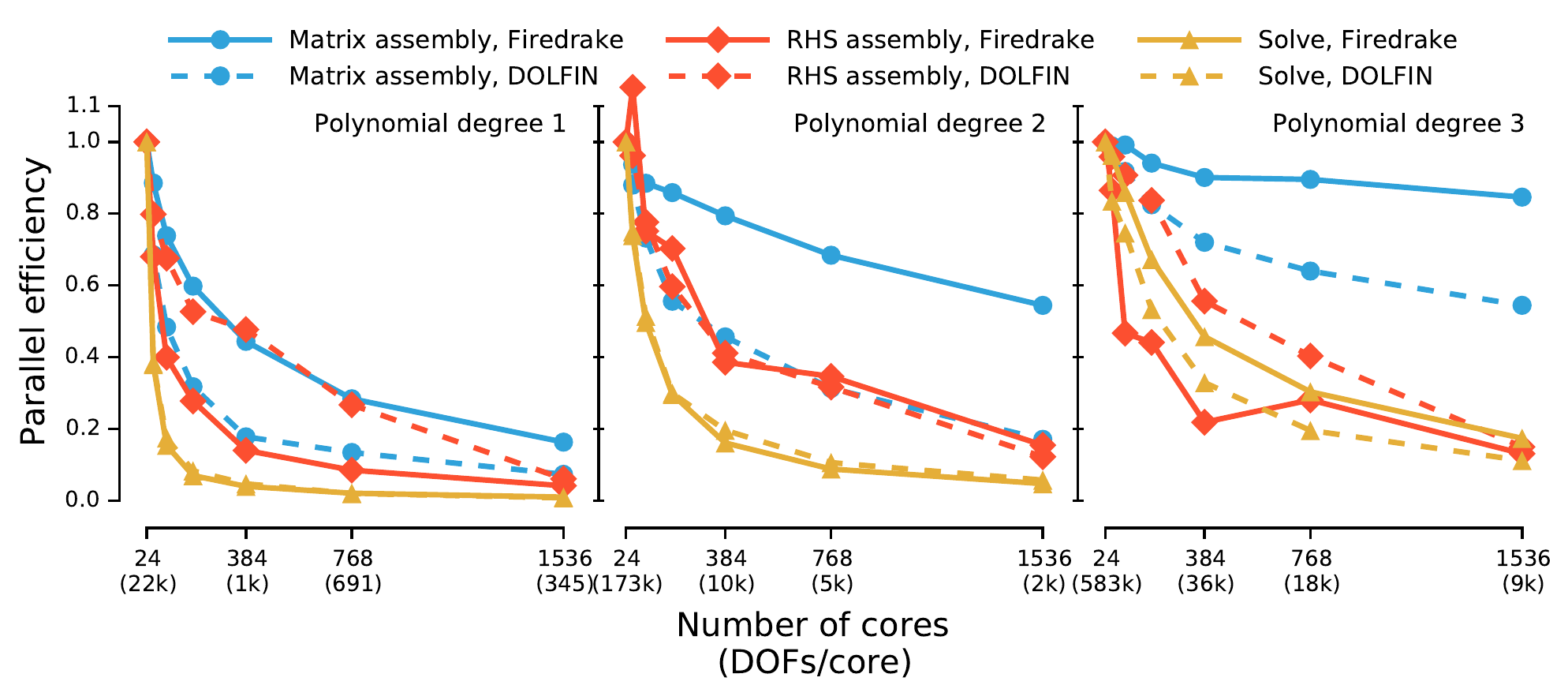}
\caption{Poisson strong scaling efficiency with respect to a full node (24
cores) on up to 1536 cores for degree one, two and three basis functions
(left to right). Firedrake matrix assembly shows the highest efficiency
across the board, whereas right-hand side assembly tails off compared to
DOLFIN due to the faster baseline performance.  Solver efficiencies are almost
identical, with a slight advantage for Firedrake at third order.}
\label{img:experiments:poisson:strong-efficiency}
\end{figure}

Parallel efficiency for the strong scaling results with respect to a full node
(24 cores) is shown in Fig.~\ref{img:experiments:poisson:strong-efficiency}.

\begin{figure}[htbp]
\centering
\includegraphics[width=\textwidth]{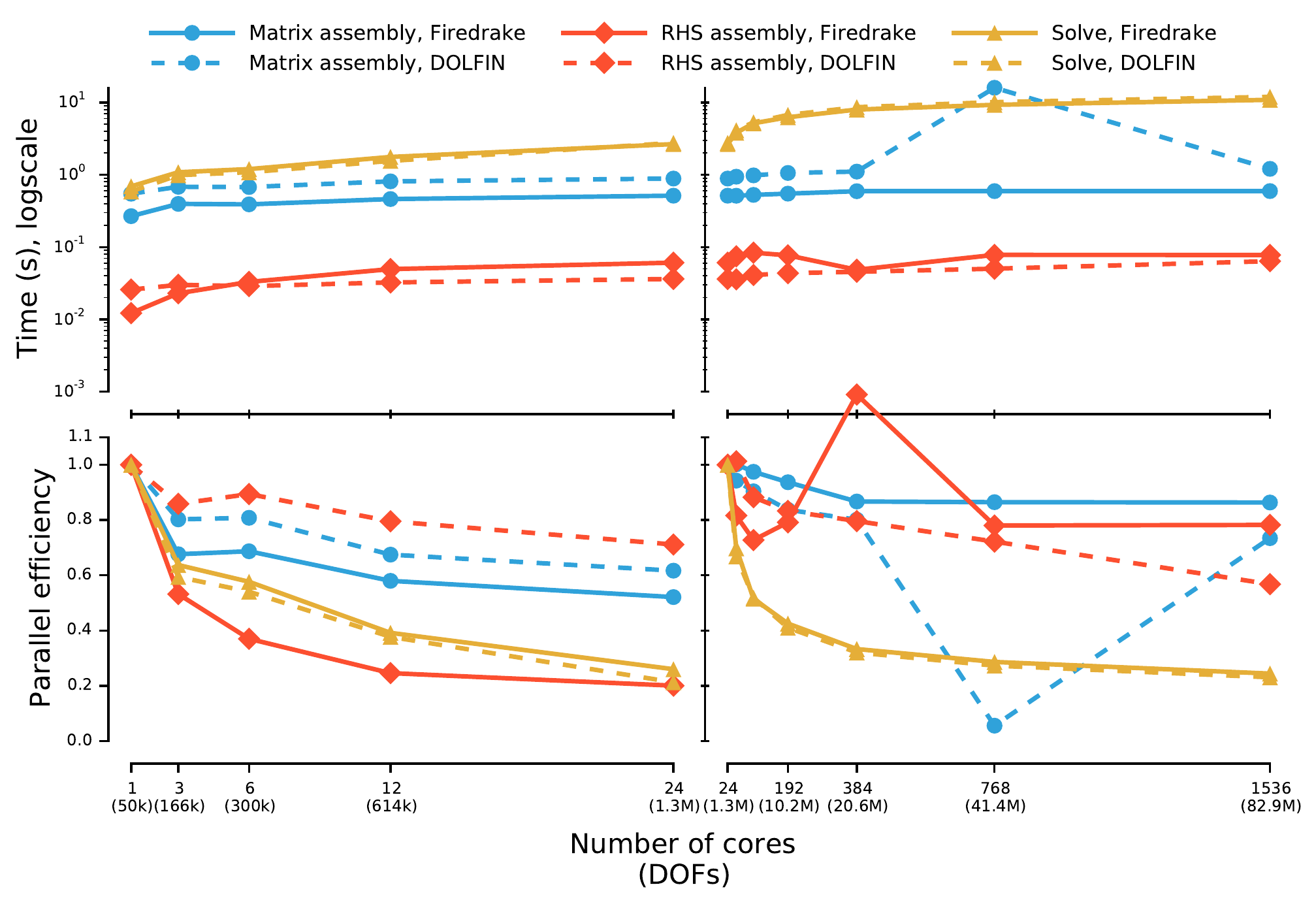}
\caption{Weak scaling performance for third order Poisson
basis functions with 50k DOFs per core.  Scaling is shown
intra-node (1-24 cores) relative to a single core (left) and
inter-node (24-1536 cores) relative to a single node (right).
Within a node, DOLFIN shows better efficiency for assembly due to Firedrake's
faster sequential baseline. In particular Firedrake right-hand side assembly
drops off significantly from one to three and three to six cores due to
resource contention, leading to DOLFIN overtaking from six cores. Beyond one
node Firedrake shows better assembly efficiency, though DOLFIN remains faster
overall for right-hand side assembly. Solver runtimes and efficiencies are
almost identical both intra and inter node.}
\label{img:experiments:poisson:weak}
\end{figure}

Weak scaling run times and efficiencies for P3 basis functions are shown in
Fig.~\ref{img:experiments:poisson:weak} separately for the intra node case
for up to 24 cores and the inter node case for 24 to 1536 cores. Within a
node, processes share resources, in particular memory bandwidth, which limits
achievable performance for these bandwidth bound computations. Scaling beyond
a node, resources per core remain constant, and the limiting factor for
scalability is network communication latency.

\subsection{Linear Wave Equation}

The strong form of the wave equation, a linear second-order PDE, is given as:
\begin{align}
 \frac{\partial^2\phi}{\partial t^2} - \nabla^2 \phi &= 0 \\
 \nabla \phi \cdot n &= 0 \ \textrm{on}\ \Gamma_N \\
 \phi &= \frac{1}{10\pi}\cos(10\pi t)  \ \textrm{on}\ \Gamma_D
\end{align}

To facilitate an explicit time stepping scheme, an auxiliary quantity $p$ is
introduced:
\begin{align}
  \label{eq:experiments:wave:phi}
 \frac{\partial\phi}{\partial t} &= - p \\
  \label{eq:experiments:wave:p}
 \frac{\partial p}{\partial t} + \nabla^2 \phi &= 0 \\
 \nabla \phi \cdot n &= 0 \ \textrm{on}\ \Gamma_N \\
 p &= \sin(10\pi t)  \ \textrm{on}\ \Gamma_D
\end{align}

The weak form of \eqref{eq:experiments:wave:p} is formed as: find $p \in V$
such that
\begin{equation}
 \int_\Omega \frac{\partial p}{\partial t} v\,\mathrm d x = \int_\Omega \nabla\phi\cdot\nabla v\,\mathrm d x
 \quad \forall v \in V
  \label{eq:experiments:wave:weak}
\end{equation}
for a suitable function space $V$. The absence of spatial derivatives in
\eqref{eq:experiments:wave:phi} makes the weak form of this equation
equivalent to the strong form so it can be solved pointwise.

An explicit symplectic method is used in time, where $p$ and $\phi$ are offset
by a half time step. Time stepping $\phi$ in \eqref{eq:experiments:wave:phi}
is a pointwise operation, whereas stepping forward $p$ in
\eqref{eq:experiments:wave:weak} involves inverting a mass matrix. However, by
lumping the mass, this operation can be turned into a pointwise one, in which the
inversion of the mass matrix is replaced by a pointwise multiplication by the
inverse of the lumped mass.

This benchmark demonstrates a numerical scheme in which no linear system is
solved and therefore no PETSc solver is invoked. The expression compiler is
used for the $p$ and $\phi$ updates and all aspects of the computation are
under the control of Firedrake. The implementation of this problem in
Firedrake is given in listing \ref{lst:wave}.

\begin{lstlisting}[language={[firedrake]python},style=framed,float,
caption={Firedrake code for the linear wave equation. The constant factor for
the $\phi$ update and the form and the constant factor $v \mathrm d x$ for the
$p$ update are precomputed such that only $\nabla\phi\cdot\nabla v\,\mathrm d
x$ is assembled each time step. The expressions \texttt{phi\_update}, and
\texttt{p\_constant} are purely symbolic, and are used
by Firedrake to generate and execute pointwise update calculations when the
\texttt{assign}, \texttt{+=}, and \texttt{-=} operations are
called. \texttt{assemble(v*dx)} is the lumped mass, an integral which is
calculated outside the form and then symbolically substituted into the
pointwise update of \texttt{p}. \texttt{p\_form} is similarly a symbolic
integral which is numerically calculated by the \texttt{assemble} call in
the \texttt{p} update. This means that the \texttt{p} update amounts to
assembling the right hand side of \eqref{eq:experiments:wave:weak} and then
using this to approximately solve \eqref{eq:experiments:wave:p} by scaling
with the timestep and multiplying (DOF by DOF) with the inverse lumped mass matrix.},label=lst:wave]
from firedrake import *
mesh = Mesh("wave_tank.msh")

V = FunctionSpace(mesh, 'Lagrange', 1)
p = Function(V, name="p")
phi = Function(V, name="phi")

u = TrialFunction(V)
v = TestFunction(V)

p_in = Constant(0.0)
bc = DirichletBC(V, p_in, 1)  # Boundary condition for y=0

T = 10.
dt = 0.001
t = 0
phi_update = dt / 2 * p
p_constant = dt / assemble(v*dx)
p_form = inner(grad(v), grad(phi))*dx
while t <= T:
    p_in.assign(sin(2*pi*5*t))
    phi -= phi_update
    p += assemble(p_form) * p_constant
    bc.apply(p)
    phi -= phi_update
    t += dt
\end{lstlisting}

\subsubsection{Results}

\begin{figure}[htbp]
\centering
\includegraphics[width=\textwidth]{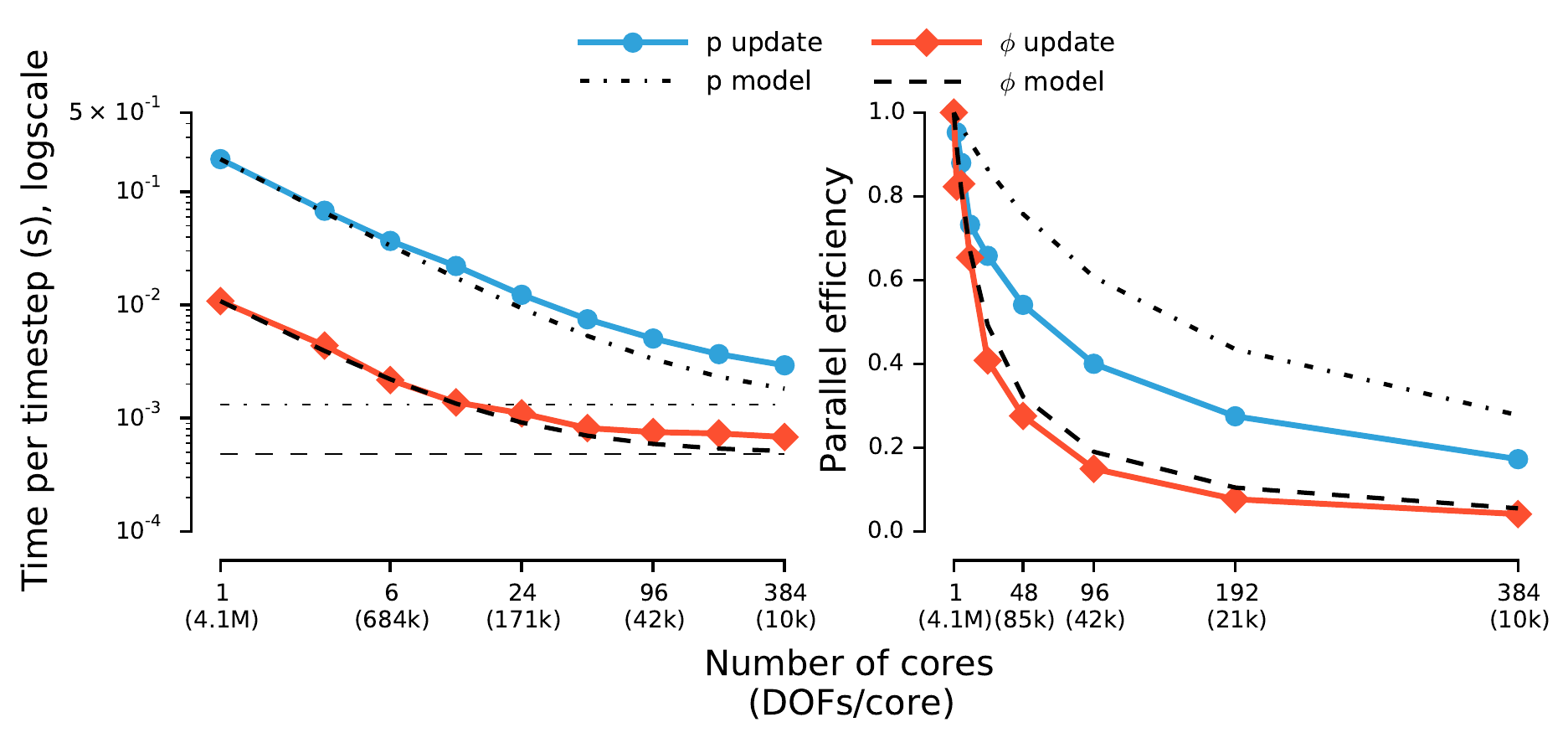}
\caption{Strong scaling (left) and parallel efficiency (right) for the $p$ and
$\phi$ updates of the explicit wave equation shown in
listing~\ref{lst:wave}.  Horizontal dashed (dotted) lines show the
non-parallelisable overhead for the $\phi$ ($p$) updates.  Given these
overheads, models for expected runtime are shown for both updates.

The $\phi$ update is a very simple expression executed as a direct loop and
follows the projected scaling curve (dashed) based on the sequential run time
and the overhead almost perfectly.  The $p$ update involves assembling a
vector, which is executed as an indirect loop and requires exchanging halo
data. Therefore, the measured scaling trails behind the projected scaling due
to communication overhead already starting at 3 cores.  Caching of the
assembled expressions in the expression compiler keeps the sequential
overheads low.}
\label{img:experiments:wave:strong-efficiency}
\end{figure}

\begin{figure}[htbp]
\centering
  \includegraphics[width=.95\textwidth]{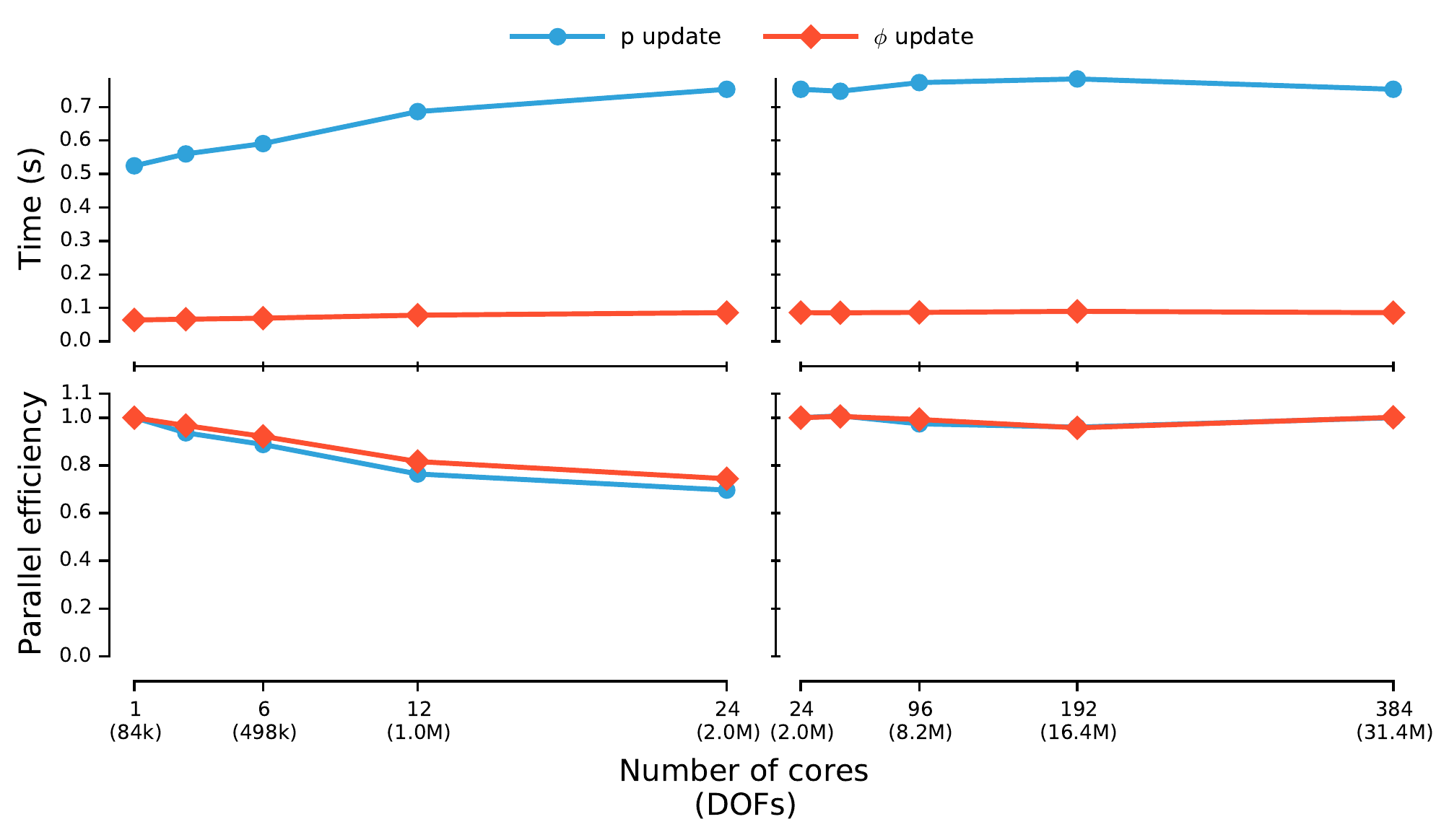}
\caption{Weak scaling performance for the explicit wave equation with
84k DOFs per core.  Scaling is shown intra-node (1-24 cores) relative
to a single core (left) and inter-node (24-384 cores) relative to a
single node (right).  Timings are for 100 timesteps.
The $\phi$ and $p$ update show a similarly high level of efficiency intra
node, only dropping to about 80\% for 24 cores. Across nodes, scaling is
almost perfect for both $\phi$ and $p$ update.}
\label{img:experiments:wave:weak}
\end{figure}

Strong scaling performance is shown in
Fig.~\ref{img:experiments:wave:strong-efficiency} for up to 384 cores and is
limited by the measured non-parallelisable overhead indicated by the
horizontal lines in the graph.  Weak scaling runtimes and efficiencies are
shown in Fig.~\ref{img:experiments:wave:weak} separately for the intra node
case for up to 24 cores and the inter node case for 24 to 384 cores.

\subsection{Cahn-Hilliard Equation}\label{sec:cahnhilliard}

The final experiment presented in this section,
is the fourth-order parabolic time-dependent non-linear Cahn-Hilliard equation,
based on a DOLFIN demo\footnote{\url{
http://fenicsproject.org/documentation/dolfin/1.6.0/python/demo/documented/cahn-hilliard/python/documentation.html}},
which involves first-order time derivatives, and second- and fourth-order
spatial derivatives. It describes the process of phase separation of the two
components of a binary fluid:
\begin{align}
   \frac{\partial c}{\partial t} - \nabla \cdot \left( M \nabla\left(\frac{d f}{d c}
             - \lambda \nabla^{2}c\right)\right) &= 0 \quad {\rm in} \ \Omega, \\
   \nabla\left(\frac{d f}{d c} - \lambda \nabla^{2}c\right)\cdot n &= 0 \quad {\rm on} \ \partial\Omega, \\
   \nabla c \cdot n &= 0 \quad {\rm on} \ \partial\Omega
  \label{eq:experiments:cahn-hilliard:strong}
\end{align}
with $c$ the unknown fluid concentration, $f$ a non-convex function in $c$,
$M$ the diffusion coefficient and $n$ the outward pointing boundary normal.

Introducing an auxiliary quantity $\mu$ (the chemical potential)
allows the equation to be restated as two coupled second-order equations:
\begin{align}
   \frac{\partial c}{\partial t} - \nabla \cdot M \nabla\mu  &= 0 \quad {\rm in} \ \Omega, \\
   \mu -  \frac{d f}{d c} + \lambda \nabla^{2}c &= 0 \quad {\rm in} \ \Omega.
  \label{eq:experiments:cahn-hilliard:mixed}
\end{align}

The time-dependent variational form of the problem with unknown fields $c$ and
$\mu$ is given as: find $(c, \mu) \in V \times V$ for a suitable function
space $V$ such that
\begin{align}
   \int_{\Omega} \frac{\partial c}{\partial t} q \, {\rm d} x +
     \int_{\Omega} M \nabla\mu \cdot \nabla q \, {\rm d} x
     &= 0 \quad \forall \ q \in V,  \\
   \int_{\Omega} \mu v \, {\rm d}x - \int_{\Omega} \frac{d f}{d c} v \, {\rm d}x
     - \int_{\Omega} \lambda \nabla c \cdot \nabla v \, {\rm d} x
     &= 0 \quad \forall \ v \in V.
  \label{eq:experiments:cahn-hilliard:weak}
\end{align}

Applying the Crank-Nicolson scheme for time discretisation yields:
\begin{align}
   \int_{\Omega} \frac{c_{n+1} - c_{n}}{dt} q \, {\rm d} x
     + \int_{\Omega} M \nabla \frac{1}{2}(\mu_{n+1} + \mu_{n}) \cdot \nabla q \, {\rm d} x
     &= 0 \quad \forall \ q \in V  \\
   \int_{\Omega} \mu_{n+1} v  \, {\rm d} x
     - \int_{\Omega} \frac{d f_{n+1}}{d c} v  \, {\rm d} x
     - \int_{\Omega} \lambda \nabla c_{n+1} \cdot \nabla v \, {\rm d} x
     &= 0 \quad \forall \ v \in V
  \label{eq:experiments:cahn-hilliard:time-discretised}
\end{align}

\subsubsection{Problem setup}

The problem is solved on the unit square, represented as a fully unstructured
mesh, with $f = 100c^2 (1-c^2)$, $\lambda = 0.01$, $M = 1$ and $dt =
5\cdot10^{-6}$. The function space $V$ is the space of first order Lagrange
basis functions.

Firedrake allows the initial condition to be set by defining a custom |Kernel|
and executing a parallel loop, in which the expression may be written as a C
string. The custom |Kernel| used to set the initial condition is shown as
Listing \ref{lst:kernel}. For comparison, an equivalent Kernel using the
lower-level PyOP2 interface is provided in Listing \ref{lst:pyop2kernel}.
\begin{lstlisting}[language={[firedrake]python},style=framed,float,
  caption={Code for a custom Firedrake kernel setting the random initial condition for
  the fluid concentration $c$ in the Cahn-Hilliard example. Inclusion of extra
  headers to use library functions and an extra piece of setup code to be
  executed only once are not usually required for custom
  kernels.},label=lst:kernel]
# Setup code setting the random seed depending on MPI rank (executed once)
setup_code = """int __rank;
MPI_Comm_rank(MPI_COMM_WORLD, &__rank);
srandom(2 + __rank);"""
# Expression setting the random initial condition
rand_init = "A[0] = 0.63 + 0.02*(0.5 - (double)random()/RAND_MAX);"
par_loop(kernel=rand_init, measure=direct, args={'A': (u[0], WRITE)},
         headers=["#include <stdlib.h>", "#include <mpi.h>"],
         user_code=setup_code)
\end{lstlisting}
\begin{lstlisting}[language={[firedrake]python},style=framed,float,
  caption={Code for a custom PyOP2 kernel equivalent to the Firedrake kernel
    in listing \ref{lst:kernel}.},label=lst:pyop2kernel]
# Setup code setting the random seed depending on MPI rank (executed once)
setup_code = """int __rank;
MPI_Comm_rank(MPI_COMM_WORLD, &__rank);
srandom(2 + __rank);"""
# PyOP2 C kernel string setting the random initial condition
rand_init = """void u_init(double A[1]) {
  A[0] = 0.63 + 0.02*(0.5 - (double)random()/RAND_MAX);
}"""
u_init = pyop2.Kernel(code=rand_init, name="u_init",
                      headers=["#include <stdlib.h>", "#include <mpi.h>"],
                      user_code=setup_code)
pyop2.par_loop(kernel=u_init, it_space=u.function_space().node_set[0],
               u.dat[0](op2.WRITE))
\end{lstlisting}

To solve the mixed system, a GMRES solver with a fieldsplit preconditioner
using a lower Schur complement factorisation 
is employed. When
solving a mixed system with a $2\times2$ block matrix with blocks $A$, $B$,
$C$, $D$ the Schur complement $S$ is given by
\begin{equation}
  S = D - C A^{-1} B.
  \label{eq:experiments:cahn-hilliard:schur-complement}
\end{equation}
and the lower factorisation is an approximation to
\begin{equation}
  \left(\begin{matrix}A & 0 \\ C & S\end{matrix}\right)^{-1} =
  \left(\begin{matrix}A^{-1} & 0 \\ 0 & S^{-1}\end{matrix}\right)
  \left(\begin{matrix}I & 0 \\ -C A^{-1} & I\end{matrix}\right).
  \label{eq:experiments:cahn-hilliard:schur-lower}
\end{equation}
where $A^{-1}$ and $S^{-1}$ are never explicitly formed.

An approximation to $A^{-1}$ is computed using a single V-cycle of the HYPRE
Boomeramg algebraic multigrid preconditioner \cite{Falgout2006}. The inverse
Schur complement, $S^{-1}$, is approximated by
\begin{equation}
  S^{-1} \approx \hat{S}^{-1} = H^{-1} M H^{-1},
\end{equation}
using a custom PETSc |mat| preconditioner, where $H$ and $M$ are defined as
\begin{align}
  H &= \sqrt{a} \langle u, v \rangle + \sqrt{c} \langle \nabla u, \nabla v \rangle
  \quad \forall v \in V \times V \\
  M &= \langle u, v \rangle \quad \forall v \in V \times V
  \label{eq:experiments:cahn-hilliard:schur-pc}
\end{align}
with $a = 1$ and $b = \frac{{\rm dt} * \lambda}{1 + 100 {\rm dt}}$
\cite{bosch2014}.

\subsubsection{Results}

\begin{figure}[htbp]
\centering
\includegraphics[width=\textwidth]{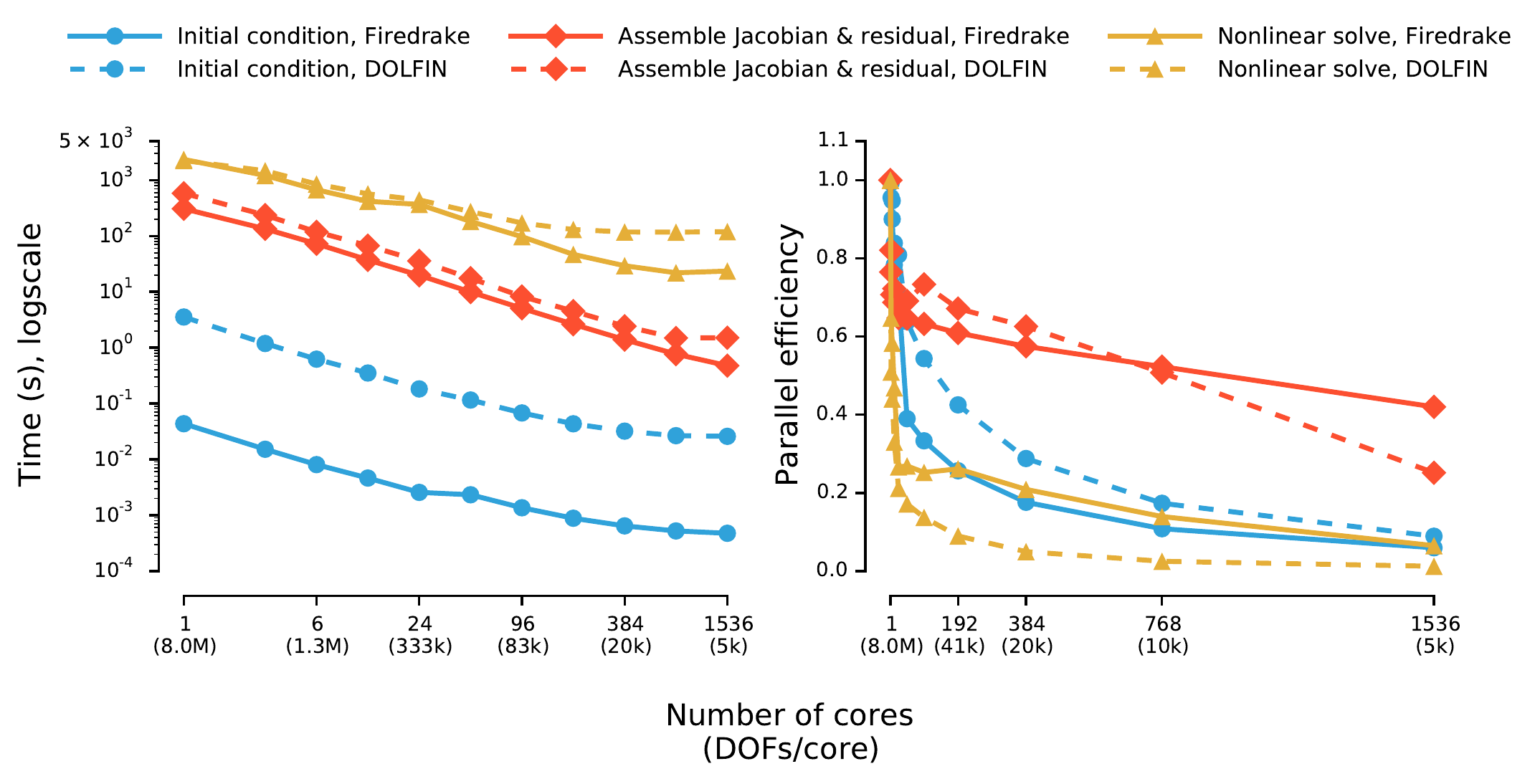}
\caption{Strong scaling (left) and parallel efficiency (right) for a
Cahn-Hilliard problem with 8M DOFs for ten time steps on up to 1536 cores.
Both Firedrake and DOLFIN achieve close to linear scaling for assembly down
to 10k DOFs per core. Firedrake is consistently faster by about a factor two,
demonstrating the efficiency of assembling mixed spaces using the form
splitting approach described in Section \ref{sec:mixed}.

Evaluating the initial condition with Firedrake is faster by about two orders
of magnitude, demonstrating the efficiency of expression evaluation using a
PyOP2 kernel as opposed to a C++ virtual function call required for DOLFIN.
Scaling flattens out in both cases from about 40k DOFs per core due to
non-parallelisable overheads.  Solver scaling is initially equivalent, with
Firedrake gaining significantly starting from about 80k DOFs per core. This is
due to the use of a PETSc MATNEST, which is more efficient when using a
fieldsplit preconditioner by avoiding expensive copies for extracting sub
blocks of the matrix.

The parallel efficiency for strong scaling shows initial advantages for DOLFIN
for assembly due to the faster sequential baseline of Firedrake, which catches
up at 10k DOFs per core. Efficiency for evaluating the initial condition shows
an advantage for DOLFIN again due to a faster Firedrake baseline and is
considerably lower than assembly due to non-parallelisable overheads.  Solver
efficiency is considerably higher for Firedrake.}
\label{img:experiments:cahn-hilliard:strong}
\end{figure}

\begin{figure}[htbp]
\centering
\includegraphics[width=.98\textwidth]{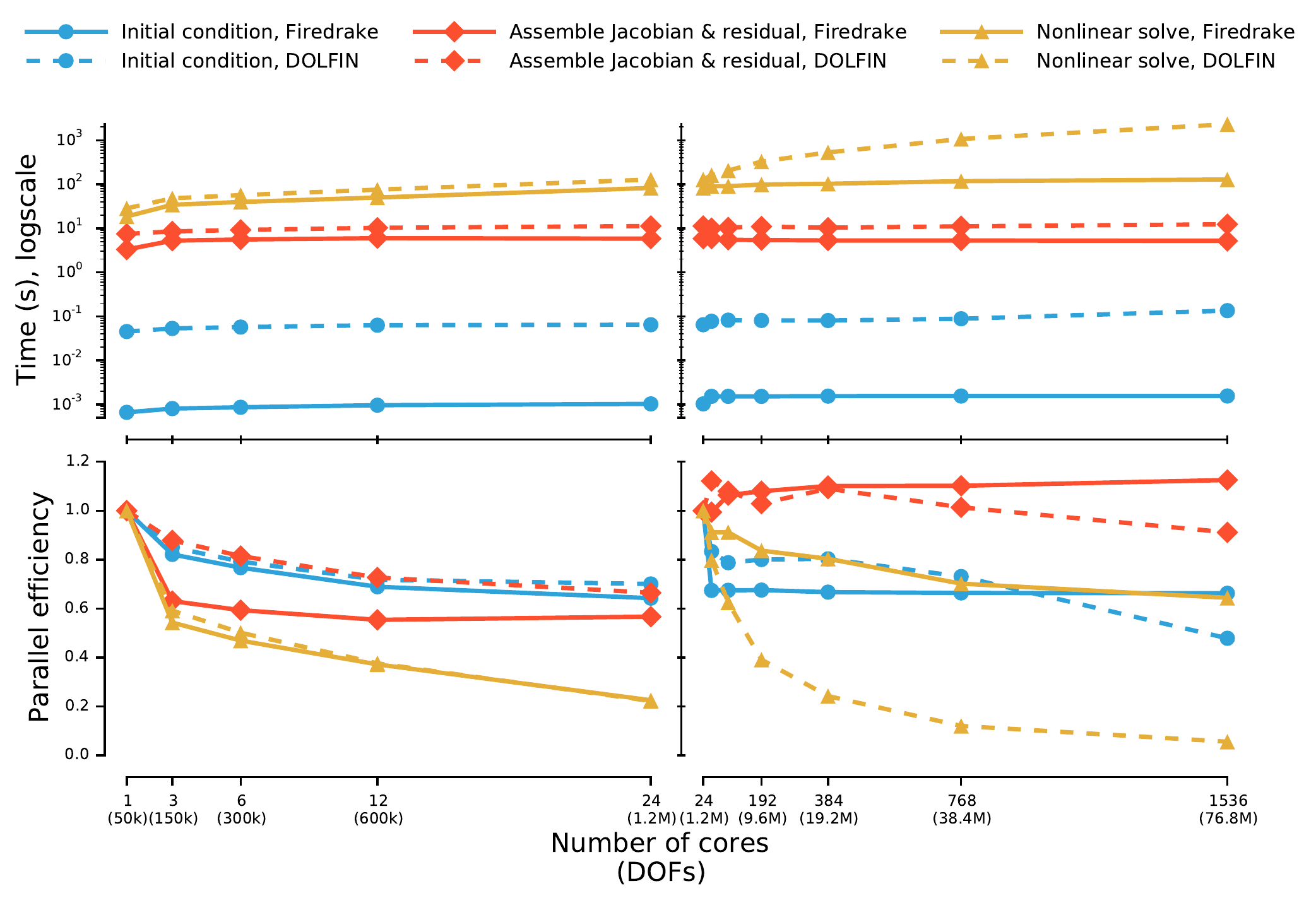}
\caption{Weak scaling performance for the Cahn-Hilliard problem with 50k DOFs
per core for 10 time steps.  Scaling is shown intra-node (1-24 cores) relative
to a single core (left) and inter-node (24-1536 cores) relative to a single
node (right).

Intra node scaling is very similar for Firedrake and DOLFIN with very good
efficiencies for all but the solve. Firedrake is faster for assembly and solve
by about a factor two and almost two orders of magnitude for the evaluation of
the initial condition.
Inter node, weak scaling for assembly is almost perfect and even superlinear for
Firedrake. Efficiency for the initial condition stabilises at just below 70\%.
The efficiency of the DOLFIN solve however slumps, which can be attributed to
memory allocations and deallocations required for building the monolithic
preconditioner, whereas Firedrake exploits the PETSc MATNEST.
}
\label{img:experiments:cahn-hilliard:weak}
\end{figure}

Strong scaling runtimes for up to 1536 cores comparing Firedrake and DOLFIN
for solving the non-linear system, assembling the residual and Jacobian forms
as well as evaluating the initial condition on an 8M DOF mesh for ten time
steps are shown in Fig.~\ref{img:experiments:cahn-hilliard:strong}.
Weak scaling run times and parallel efficiencies are shown separately for 1-24
cores intra and 24-1536 cores inter node in Fig.~\ref{img:experiments:cahn-hilliard:weak}.

\subsection{Performance discussion}

The experiments presented were selected to demonstrate the performance of
Firedrake in several different regimes. By drawing together the results, we
can make some observations on the impact of the introduction of the PyOP2
abstraction layer and its implementation.

\subsubsection{Assembly in comparison with DOLFIN}

First, assembly of linear and bilinear forms in Firedrake is consistently
much faster than in DOLFIN. There are several features of Firedrake which
impact on this. Critically, the PyOP2 interface is an abstract basis for
code generation, while the UFC interface imposed by DOLFIN is a C++ abstract
interface \cite{Alnaes2012}. This means that PyOP2 kernels can be
completely inlined while DOLFIN kernels result in multiple virtual function
calls per element. The COFFEE optimisations have been found to result in up
to a fourfold increase in speed over the quadrature optimisations in
FFC \cite{Luporini2015}. The speedup is most pronounced in the case of
the Cahn-Hilliard equation, which employs mixed function spaces. In this
case, a performance increase is expected due to the form splitting
optimisation (see Section \ref{sec:mixed}).

\subsubsection{Scaling performance}

The weak scaling performance of pure Firedrake code (that is, excluding the
PETSc solver) beyond one node is uniformly excellent. Within one node,
resource contention results in significantly less than perfect efficiency, but
this is expected. In the strong scaling regime, the fixed overhead per field
of some hundreds of microseconds (Fig.~\ref{img:experiments:cahn-hilliard:strong}) results in loss of optimal
scaling at a significantly higher degree of freedom count than would be
completely optimal. Reduction of the fixed overhead therefore remains an
important development objective.

\section{Current limitations and future extensions}

\subsection{Accelerators and threads}\label{sec:nogpu}

This paper presents only performance results for MPI parallel execution on
CPUs, with instruction level vector parallelism facilitated by COFFEE. As Fig.~\ref{fig:firedrake_toolchain} shows, PyOP2 also supports execution using
OpenMP threads or OpenCL on CPU, and OpenCL and CUDA on GPU. Preliminary
performance results on these platforms were published in \citeN{Markall2013}.
However, the available hybrid parallel and GPU linear solver libraries are far
more limited than PETSc's MPI-only functionality. The Firedrake developers
have therefore given priority to achieving high performance and feature
completeness for the CPU backend using MPI and vector parallelism. The other
backends are fully functional in the sense that form assembly is supported and
solving is supported to the limits of the relevant solver backends. This
demonstrates the utility of the PyOP2 interface in isolating such
implementation matters from the specification of the algorithm. However at
this stage only the MPI CPU backend is considered to be of production quality
and suitable for full documentation here. Given the increasingly fine-grained
parallelism of both CPU and accelerator hardware, hybrid parallel approaches
combining message passing with shared memory approaches will be a future
direction of development.

\subsection{$hp$-adaptive finite element methods}

Support for $p$-refined finite element methods requires lifting the
restriction of PyOP2 maps to fixed arity described in Section
\ref{sec:pyop2:concepts:maps}. Permitting variable arity maps and the
consequent variable trip count loops in kernels would impede many of the
low-level optimisations applied by COFFEE such that both classes of maps
should be supported independently. The map storage format would also be
required to record the arity of each source element. A more promising option
would be to support container maps containing several maps of different arity
and corresponding kernels to match. This would enable the support of not just
$p$-refinement, but also mixed geometry meshes.

\subsection{Firedrake-adjoint}

\citeN{Farrell2012a}\ demonstrated that the mathematical abstraction
captured by the FEniCS language can be exploited to automate the generation
and highly efficient execution of the tangent linear and adjoint models
corresponding to forward models written in that
language. Dolfin-adjoint\footnote{\url{http://dolfin-adjoint.org}}, the
software implementing \citeN{Farrell2012a}, operates on objects at the
FEniCS Language level. Using only a short Python wrapper module,
dolfin-adjoint has been extended to support Firedrake solvers written using
unextended versions of the FEniCS Language. The user-defined extension
kernels described in section \ref{sec:escape}\ are not supported by this
Firedrake-adjoint, since they cannot be differentiated using UFL's intrinsic
symbolic operations. The extension of Firedrake-adjoint to employ
traditional algorithmic differentiation methods to custom kernels is planned
for the future.

\section{Acknowledgements}

The authors would like to thank Patrick E.~Farrell for his help with the
Cahn-Hilliard preconditioner implementation and helpful comments on the
results section. We would further like to thank the other contributors whose
code is in Firedrake: Mikl\'os Homolya, George Boutsioukis, Nicolas Loriant, and Kaho
Sato. Finally we would like to acknowledge the input and ideas we receive
from Colin J.~Cotter and from the core FEniCS development team, particularly
Martin S.~Aln\ae s and Marie E.~Rognes.

\bibliographystyle{ACM-Reference-Format-Journals}
\bibliography{bibliography}
\end{document}